\documentstyle[epsfig,amsmath,amssymb,12pt]{article}
\begin{document}
\setlength{\parskip}{0.45cm}
\setlength{\baselineskip}{0.75cm}
%
%
%
\begin{titlepage}
\setlength{\parskip}{0.25cm}
\setlength{\baselineskip}{0.25cm}
\begin{flushright}
DO-TH 2000/15\\
\vspace{0.2cm}
hep--ph/0011300\\
\vspace{0.2cm}
November 2000
\end{flushright}
\vspace{1.0cm}
\begin{center}
\LARGE
{\bf On the Determination of the Polarized Sea
 Distributions of the Nucleon}
\vspace{1.5cm}

\large
M. Gl\"uck, A.\ Hartl, and E.\ Reya\\
\vspace{1.0cm}

\normalsize
{\it Universit\"{a}t Dortmund, Institut f\"{u}r Physik,}\\
{\it D-44221 Dortmund, Germany} \\
\vspace{0.5cm}

\vspace{1.5cm}
\end{center}

\begin{abstract}
\noindent
The possibilities to determine the flavor structure of the polarized
sea (antiquark) distributions of the nucleon via vector boson
$(\gamma^*,\ W^{\pm},\, Z^0)$ production at high energy polarized
hadron--hadron colliders, such as the Relativistic Heavy--Ion Collider
(RHIC), are studied in detail.  In particular the perturbative stability
of the expected asymmetries in two representative models for the
(un)broken flavor structure are investigated by confronting perturbative
QCD leading order predictions of the expected asymmetries with their
next--to--leading order counterparts.
\end{abstract}
\end{titlepage}

\renewcommand{\theequation}{\arabic{section}.\arabic{equation}}
\section{Introduction}

The determination of the polarized parton content of the nucleon via
measurements of the inclusive structure functions $g_1^{p,n}(x,Q^2)$ does
not provide detailed information concerning the flavor structure of these
distributions,  in
complete analogy to unpolarized deep inelastic structure functions.  In
particular the flavor structure of the antiquark (sea) distributions is
not fixed and one needs to resort to semi--inclusive deep inelastic hadron
production for this purpose  \cite{ref1,ref2}.
The resulting antiquark distributions $\Delta\bar{q}$ are, however, not
reliably determined by this method \cite{ref3,ref4} for the time being
due to their dependence on the rather poorly known quark fragmentation
functions at low scales.

A more reliable determination is provided via inclusive vector boson
$(\gamma^*,\, W^{\pm},\, Z^0)$ production in polarized hadron--hadron
collisions as envisaged at RHIC (BNL) or at HERA-$\vec{\rm N}$ (DESY)
whose potential for the determination of $\Delta\bar{u}$ and $\Delta\bar{d}$
will be studied in this paper in leading (LO) {\em{as well as}}
next--to--leading (NLO) order of QCD.

Recently the LO Drell--Yan ($\gamma^*$) dilepton production process for
future polarized $\vec{p}\vec{p}$ and $\vec{p}\,\vec{d}$ collision
experiments
has been suggested and studied \cite{ref5,ref6} for delineating the
flavor--asymmetry of the polarized light sea distributions $\Delta\bar{u}
(x,Q^2)\neq \Delta\bar{d}(x,Q^2)$.  In particular, the relativistic field
theoretic chiral quark--soliton model \cite{ref7,ref8,ref9,ref2} as well
as a more recent analysis based on the statistical parton model \cite{ref10}
predict $|\Delta\bar{d}(x,Q^2)|>\Delta\bar{u}(x,Q^2)$.  Similar expectations
\cite{ref11,ref7,ref12,ref10,ref13}, partly from first principles, hold also
for the unpolarized light sea distributions, $\bar{d}>\bar{u}$, which have
been already confirmed, as is well known, by Drell--Yan $\mu^+\mu^-$
production \cite{ref14} $pp$ and $pd$ experiments \cite{ref15}.  In general,
the flavor--asymmetry of the light sea distributions can be understood
in terms of flavor mass asymmetries and `Pauli--blocking' effects being
related to the Pauli exclusion principle \cite{ref16,ref17,ref13}.

In order to analyze the determination of (flavor--asymmetric) light
antiquark
(sea) distributions one needs some alternative models for $\Delta\bar{q}$,
introduced in Section 2, and the possibility of their experimental
distinction
will be investigated in Section 3. Our conclusions are drawn in Sect.\ 4
and the Appendix contains all expressions of LO and NLO cross sections
relevant for our calculations.

\section{Models for the Polarized Antiquark Distributions of the Nucleon}

(i)  Standard unbroken sea scenario: Here one assumes,  as in most analyses
of polarization data performed thus far, a flavor symmetric sea, i.e.\,
\begin{equation}
\Delta\bar{u}(x,Q^2)=\Delta\bar{d}(x,Q^2)=\Delta\bar{s}(x,Q^2)\equiv
\Delta\bar{q}(x,Q^2)
\end{equation}
where as usual $\Delta u_{\rm sea}=\Delta\bar{u}$,  $\Delta d_{\rm sea}
=\Delta\bar{d}$ and $\Delta s=\Delta\bar{s}$.
The adopted LO and NLO distributions will be taken
from the recent analysis of the AA Collaboration \cite{ref18}, in
particular their LO and NLO--1 ones.

\noindent (ii) Broken sea scenario:  Here one assumes $\Delta\bar{u}
(x,Q^2)\neq \Delta\bar{d}(x,Q^2)\neq \Delta\bar{s}(x,Q^2)$, i.e.\ a
broken flavor symmetry as motivated by the situation in the corresponding
unpolarized sector \cite{ref15,ref13}.  As mentioned in the Introduction,
present polarization data do not provide detailed and reliable information
concerning flavor symmetry breaking and we shall therefore utilize
antiquark distributions extracted via the phenomenological ansatz of
\cite{ref17} which is confirmed in the unpolarized sector and which
moreover agrees well with the predictions obtained within the framework
of the chiral quark--soliton model \cite{ref7,ref8,ref9,ref2} and with
a recent analysis of semi--inclusive deep inelastic data \cite{ref4}.

These flavor--asymmetric distributions, henceforth denoted by $\Delta f'$,
are related to the (flavor--symmetric) AAC \cite{ref18} distributions
$\Delta f$ by the relations

\begin{eqnarray}
\Delta u'(x,Q_0^2) & = & \Delta u(x,Q_0^2)-\phi(x)\nonumber\\
\Delta\bar{u}'(x,Q_0^2) & = & \Delta\bar{q}(x,Q_0^2)+\phi(x)\nonumber\\
\Delta d'(x,Q_0^2) & = & \Delta d(x,Q_0^2)+\phi(x)\nonumber\\
\Delta\bar{d}'(x,Q_0^2) & = & \Delta\bar{q}(x,Q_0^2)-\phi(x)\nonumber\\
\Delta u'_v(x,Q_0^2) & = & \Delta u_v(x,Q_0^2)-2\phi(x)\nonumber\\
\Delta d'_v(x,Q_0^2) & = & \Delta d_v(x,Q_0^2)+2\phi(x)\nonumber\\
\Delta s'(x,Q_0^2) & = & \Delta\bar{s}'(x,Q_0^2)=\Delta s(x,Q_0^2)=
  \Delta\bar{s}(x,Q_0^2)=\Delta\bar{q}(x,Q_0^2)\nonumber\\
\Delta g'(x,Q_0^2) & = & \Delta g(x,Q_0^2)
\end{eqnarray}
with $\Delta q_v\equiv\Delta q-\Delta\bar{q}$ and
$\Delta\bar{q}$ given by (2.1) at the input scale $Q_0^2=1$ GeV$^2$,
and where
\begin{equation}
\phi = -\Delta\bar{q}\,
          \frac{\Delta u -\Delta d}{\Delta u+\Delta d -2\Delta\bar{q}}
\end{equation}
as follows from the Pauli--blocking relation \cite{ref17}
\begin{equation}
\Delta\bar{d}'(x,Q_0^2)/\Delta\bar{u}'(x,Q_0^2) = \Delta u'(x,Q_0^2)/
    \Delta d'(x,Q_0^2)
\end{equation}
combined with the constraints ($q=u,\, d$)
\begin{eqnarray}
\Delta q'(x,Q_0^2) + \Delta\bar{q}'(x,Q_0^2) & = & \Delta q(x,Q_0^2)+
         \Delta\bar{q}(x,Q_0^2)\nonumber\\
\sum_{q=u,\, d} \Delta q'_v(x,Q_0^2) & =  & \sum_{q=u,\, d}\Delta
q_v(x,Q_0^2)
\end{eqnarray}
needed \cite{ref19} to preserve the quality of the fit \cite{ref18} to
$g_1^{p,n}(x,Q^2)$ within the standard unbroken sea scenario.  Note that
$\Delta\bar{u}'-\Delta\bar{d}'=2\phi$ and that the `breaking' function
$\phi(x)$ in (2.3) can be simply parametrized in LO and NLO as well,
\begin{eqnarray}
x\phi(x)_{\rm LO} & = & x^{0.536} (0.27-0.14x+13.57x^2-10.17x^3)
     (1-x)^{10.5}\nonumber\\
x\phi(x)_{\rm NLO} & = & x^{0.32}(0.12+0.90x-6.68x^2 +20.06x^3-20.36x^4)
  (1-x)^{7.7}\,\, ,
\end{eqnarray}
which allows to form Mellin $n$--moments of $\phi$ needed for performing
the $Q^2$--evolutions in $n$--moment space as described, for example, in
\cite{ref20}.  The resulting distributions are shown in fig.\ 1 at the
input scale $Q_0^2$ as well as at $Q^2=25$ GeV$^2$ and $M_W^2$, and for
comparison we also show the flavor--symmetric sea density $\Delta\bar{q}$
in (2.1) of AAC \cite{ref18}.  It
should be emphasized that our broken distributions should {\em{not}}
be used well below $x\simeq 10^{-2}$ since our breaking--ansatz
\cite{ref17,ref19} $\Delta\bar{d}'/\Delta\bar{u}'\equiv (\Delta\bar{q}-\phi)
/(\Delta\bar{q}+\phi)= (\Delta u-\phi)/(\Delta d+\phi)$ which results in
(2.3) gives rise to artificial oscillations \cite{ref19} below $x=0.01$
due to the appearance of differences of parton distributions in (2.3).
All our subsequent analyses employ these broken distributions well above
$x=10^{-2}$. (For obvious reasons we skip from now on
the  `prime' notation in the above equations for the flavor--asymmetric
distributions and simply refer to `AAC' in connection with the original
AAC--densities with their flavor--symmetric sea $\Delta\bar{q}$.)

Furthermore, the unpolarized cross sections needed for the asymmetry
calculations of Sect.\ 3 will be evaluated with the LO and NLO unpolarized
parton distributions GRV98 \cite{ref21} utilized by the AAC \cite{ref18}.

\section{Cross Section Asymmetries for Inclusive Hadronic Vector Boson
         Production}
\setcounter{equation}{0}

As stated in the Introduction, the inclusive hadronic vector boson
production yields reliable information on the antiquark (sea) content of
the nucleon.  We shall consider double (single) asymmetries as obtained
with doubly (singly) polarized hadron beams.  The relevant expressions
for the corresponding cross sections in LO and NLO are collected, for
convenience, in the Appendix.

We shall successively study $\gamma^*$ and $W^{\pm},\, Z^0$ production
in Sections 3.1 and 3.2, respectively, starting with estimates of expected
production rates at RHIC (BNL) and turning then to a study of the optimal
ways to gain insight about the flavor structure of the antiquark
distributions
in the nucleon.

\subsection{$\gamma^*$ Production}

The expected production rates at RHIC are determined via the following
expected energies and integrated luminosities \cite{ref22}:
$\sqrt{S}=50-500$ GeV, ${\cal{L}}_{\vec{p}\,\vec{p}}\,(\sqrt{S})
\simeq(\sqrt{S}/500$ GeV) $800\, pb^{-1}\,$.
The polarization rate for the proton beams is expected to be
$P_{\vec{p}}\simeq 0.7$ and will be obviously lower for polarized neutrons,
i.e.\ deuteron beams for example.  In our estimates for the expected
statistical errors we shall use $P=0.7$ everywhere which represents an
idealized underestimate of these rather crucial errors for the case of
$d(n)$ beams.

Asymmetries involving only polarized protons, although easiest to be
measured experimentally, are unlikely to provide unique signatures for
a flavor--broken polarized light sea:
\vspace{-0.3cm}
{\setlength{\jot}{0.4cm}
\begin{eqnarray}
A_{\vec{p}\,\vec{p}}^{\gamma^*} & \equiv &
   \sigma_{\vec{p}\,\vec{p}}^{\gamma^*}\, / \,\sigma_{pp}^{\gamma^*}\\
& \stackrel{\rm LO}{\simeq} & - \frac{ 4\left[ \Delta u(1)\Delta\bar{u}(2)
  +\Delta\bar{u}(1)\Delta u(2)\right] + \Delta{d}(1)\Delta\bar{d}(2) +
     \Delta\bar{d}(1)\Delta d(2)}
   {4\left[ u(1)\bar{u}(2) +\bar{u}(1)u(2)\right] + d(1)\bar{d}(2) +
       \bar{d}(1)d(2)}\nonumber
\end{eqnarray}}
\vspace{-0.5cm}

\noindent where $(\Delta)\sigma$ denotes the relevant differential cross
sections
summarized in the Appendix and where we make use of the simple abbreviated
`1,2' notation introduced in the Appendix as well.  The size and even the
sign of this asymmetry depends already strongly on the choice of the
particular set of unbroken polarized parton densities \cite{ref23}.  For
our sets of (un)broken distributions in Sect.\ 2 we obtain, for example,
at $\sqrt{S}=50$ GeV and for an invariant dilepton mass $M=5$ GeV in
$\vec{p}\,\vec{p}\to\gamma^*(M)X\to\mu^+\mu^-X$ an asymmetry
$A_{\vec{p}\,\vec{p}}^{\gamma^*}\simeq +2\%$ for the unbroken (AAC)
densities and $A_{\vec{p}\,\vec{p}}^{\gamma^*}\simeq -(4\, {\rm to}\, 6)\%$
for the flavor--broken scenario for all values of $x_F$.

As in the case of flavor--broken unpolarized parton distributions
\cite{ref14,ref15}, additional polarized $pn(pd)$ reactions are required
for delineating the flavor structure of the polarized light (anti)quark
sea $\Delta\bar{u}$ and $\Delta\bar{d}$:
{\setlength{\jot}{0.4cm}
\begin{eqnarray}
A_{\vec{p}\,\pm\,\vec{n}}^{\gamma^*} & \equiv &
   \frac{\Delta\sigma_{\vec{p}\,\vec{p}}^{\gamma^*}\, \pm \,
     \Delta\sigma_{\vec{p}\,\vec{n}}^{\gamma^*}}
   {\sigma_{pp}^{\gamma^*}+\sigma_{pn}^{\gamma^*}}\\
& \stackrel{\rm LO}{\simeq} & - \frac{ [4 \Delta u(1)\pm\Delta d(1) ]\,
   [\Delta\bar{u}(2)\pm\Delta\bar{d}(2)] + [4\Delta\bar{u}(1)
    \pm\Delta\bar{d}(1)] \, [\Delta u(2)\pm \Delta d(2)]}
   {[4 u(1)+ d(1)]\, [\bar{u}(2) +\bar{d}(2)]+
   [4\bar{u}(1)+\bar{d}(1)]\, [u(2)+ d(2)] } \nonumber\\
& \stackrel{x_F\to 1}{\longrightarrow} & - \frac{[4\Delta u(1)\pm \Delta
d(1)]\,
   [\Delta\bar{u}(2)\pm \Delta\bar{d}(2)]}
     {[4u(1)+d(1)]\, [\bar{u}(2)+\bar{d}(2)]} \nonumber
\end{eqnarray}}
{\setlength{\jot}{0.4cm}
\begin{eqnarray}
R_{p\pm n} & \equiv &
   \frac{1}{2}\left( 1 \pm \Delta\sigma_{\vec{p}\,\vec{n}}^{\gamma^*}\, /\,
     \Delta\sigma_{\vec{p}\,\vec{p}}^{\gamma^*}\right)\\
& \stackrel{\rm LO}{\simeq} &  \frac{1}{2}\,\, \frac{ [4 \Delta u(1)
     \pm\Delta d(1) ]\,
   [\Delta\bar{u}(2)\pm\Delta\bar{d}(2)] + [4\Delta\bar{u}(1)
    \pm\Delta\bar{d}(1)] \, [\Delta u(2)\pm \Delta d(2)]}
   {4 [\Delta u(1)\Delta\bar{u}(2)+\Delta\bar{u}(1) \Delta u(2)]+
   \Delta d(1)\Delta\bar{d}(2)+\Delta\bar{d}(1)\Delta d(2)} \nonumber\\
& \stackrel{x_F\to 1}{\longrightarrow} & \frac{1}{2}\,\, \frac{[4\Delta
u(1)\pm
      \Delta d(1)]\, [\Delta\bar{u}(2)\pm \Delta\bar{d}(2)]}
     {4\Delta u(1)\Delta\bar{u}(2) +\Delta d(1) \Delta\bar{d}(2)}
     \simeq \frac{1}{2} \left[ 1\pm\, \frac{\Delta\bar{d}(2)}
       {\Delta\bar{u}(2)} \right] \nonumber
\end{eqnarray}}
\vspace{-0.5cm}

\noindent where for $x_F\to1$, i.e.\ $x_1\to1$ and $x_2\to 0$, the small
$\Delta\bar{q}(1)$ terms can be neglected and in the very last `crude'
(LO) approximation of $R_{p\pm n}$ the $\Delta d(1)$ terms are neglected
with respect to  $4\Delta u(1)$.  This latter ratio of polarized cross
sections in (3.3) has been suggested and studied in LO originally in
\cite{ref5} as a very sensitive observable for testing the flavor--asymmetry
of the polarized light sea.  Indeed, at small $x$, $R_{p\pm n}$ is
sensitive directly to the ratio $\Delta\bar{d}/\Delta\bar{u}$ whereas
$A_{\vec{p}\,\pm\,\vec{n}}^{\gamma^*}$ in (3.2) is proportional to the
 isoscalar and isovector combinations $\Delta\bar{u}\pm\Delta\bar{d}$.

First we present in fig.\ 2 the unpolarized cross section
$d\sigma_{pp}^{\gamma^*}/dM\, dx_F$ for $\sqrt{S}=50$ and 100 GeV.  It is
obvious that only for
$|x_F|$ \raisebox{-0.1cm}{$\stackrel{<}{\sim}$} 0.5
and not too large dilepton masses $M$ useful production rates can be
obtained.  From
\begin{equation}
x_{1,2}^0 = \frac{1}{2}\, \left( \sqrt{x_F^2+4M^2/S} \pm x_F\right)
\end{equation}
one infers that the small--$x_2$ region, relevant for our study of
sea (antiquark) distributions, implies small $\tau \equiv M^2/S$ values.
Here, furthermore, the NLO contributions are {\em{genuine}}
${\cal{O}}(\alpha_s)$ corrections and small.  Since the relevant asymmetries
decrease with $\sqrt{S}$ we have chosen $\sqrt{S}=50$ GeV and $M$ = 5 -- 8
GeV to be the appropriate range for our study of asymmetries related to
Drell--Yan dilepton pairs notwithstanding the fact, shown in \mbox{fig.\ 2,}
 that
the unpolarized cross sections do increase with $\sqrt{S}$.

As expected, $A_{\vec{p}-\vec{n}}^{\gamma^*}$ and in particular
$R_{pd}\equiv R_{p+n}$ are the best indicators for the flavor structure
of the antiquark (sea) distibutions:  Fig.\ 3 presents our LO and NLO
results with the latter ones being, furthermore, remarkably stable with
respect to sizeable variations of the factorization scale $\mu_F$.  It
should be noted that for {\underline{all}} scenarios of polarized parton
distributions with a flavor--symmetric light sea,
$\Delta\bar{u}=\Delta\bar{d}$, we have $R_{pd}\to 1$ and
$A_{\vec{p}-\vec{n}}\to 0$ as $x_F\to 1$ -- a limit which is already
reached for
$x_F$\raisebox{-0.1cm}{$\stackrel{>}{\sim}$} 0.2 for most sets of
polarized parton densities as illustrated for AAC in fig.\ 3.  The
statistical errors shown in fig.\ 3 are obtained from
(see, e.g., \cite{ref22,ref24})
\begin{eqnarray}
\Delta A_{\vec{h}_1\,\vec{h}_2}&  \simeq  & \pm \, \frac{1}{P_1P_2}\,\,
      \frac{1}{\sqrt{4{\cal{L}}\sigma_{h_1\, h_2}}}\\
\Delta R_{p\pm n} & \simeq & \pm \, \frac{1}{2P^2}\,\,
    \frac{1}{\sqrt{(A_{\vec{p}\,\vec{p}})^2}}\,\,
           \frac{1}{\sqrt{4{\cal{L}}\sigma_{pp}}}\,\, ,
\end{eqnarray}
\vspace{-0.8cm}

\noindent assuming $P_1=P_2=P=0.7$ for the beam polarizations,
${\cal{L}}=80\,\,pb^{-1}$
and bin widths $\Delta x_F=0.1,\,\, \Delta M=1$ GeV for calculating
bin-integrated cross sections.

\subsection{$W^{\pm}$ and $Z^0$ Production}

The production of these vector bosons via (un)polarized $pp(n)$
collisions, $pp(n)\to W^{\pm}X\to \mu^{\pm}\stackrel{(-)}{\nu}_{\mu}X$
and $pp(n)\to Z^0 X\to\mu^+\mu^-X$, affords of course higher c.m.\
energies which are, however, available at RHIC ($\sqrt{S}_{\rm max}=500$
GeV) \cite{ref22}.  The expected cross sections, presented in fig.\ 4,
are comparable to those for $\gamma^*$ production at $\sqrt{S}=50$ GeV
in fig.\ 2.  These cross sections are partly more sensitive to the flavor
structure of the light sea, although at far larger scales $\mu_F\sim
M_{W^{\pm},Z^0}$, and may discern its polarized flavor structure not
only in doubly polarized collisions but also in the single polarization
mode \cite{ref25}--\cite{ref27} where only one beam is polarized.  Possible
benefits of this latter mode are the expected lower statistical errors
\begin{equation}
\Delta A_{\vec{h}_1\, h_2}\simeq \pm\, \frac{1}{P_1}\,\,
    \frac{1}{\sqrt{4{\cal{L}}\sigma_{h_1\, h_2}}}
\end{equation}
i.e., a factor $P_1^{-1}$, as compared to $(P_1P_2)^{-1}$ for the doubly
polarized mode in (3.5), as well as possibly higher luminosities.  As
seen in fig.\ 4, the relevant rapidity range is
\mbox{$|y| $\raisebox{-0.1cm}{$\stackrel{<}{\sim}$} 1}
which covers the interesting range of Bjorken--$x$ for the sea
distributions,
\mbox{0.06  \raisebox{-0.1cm}{$\stackrel{<}{\sim}$} $x$
        \raisebox{-0.1cm}{$\stackrel{<}{\sim}$} 0.4\, .}

The relevant single and double helicity asymmetries are
{\setlength{\jot}{0.4cm}
\begin{eqnarray}
A_{\vec{p}\,p}^{W^+} & \equiv & \Delta\sigma_{\vec{p}\,p}^{W^+}\, /\,
     \sigma_{pp}^{W^+}\\
& \stackrel{\rm LO}{\simeq} & \frac{-\Delta u(1)\bar{d}(2)
     +\Delta\bar{d}(1)u(2)} {u(1)\bar{d}(2)+\bar{d}(1)u(2)}\quad
  \xrightarrow\quad
    \left\{ \begin{array}{cc}
        -\Delta u(1)/u(1)\,, & y \gtrsim +\frac{1}{2}\\
         \quad\Delta\bar{d}(1)/\bar{d}(1)\,, & y\lesssim -\frac{1}{2}
          \end{array}\right.\nonumber\\
\vspace{-0.5cm}
A_{\vec{p}\,p}^{W^-} & \equiv & \Delta\sigma_{\vec{p}\,p}^{W^-}\, /\,
     \sigma_{pp}^{W^-}\\
& \stackrel{\rm LO}{\simeq} & \frac{-\Delta d(1)\bar{u}(2)
     +\Delta\bar{u}(1)d(2)} {d(1)\bar{u}(2)+\bar{u}(1)d(2)}\quad
  \xrightarrow\quad
    \left\{ \begin{array}{cc}
        -\Delta d(1)/d(1)\,, & y \gtrsim +\frac{1}{2}\\
         \quad\Delta\bar{u}(1)/\bar{u}(1)\,, & y\lesssim -\frac{1}{2}
          \end{array}\right.\nonumber\\
\vspace{-0.5cm}
A_{\vec{p}\,\vec{p}}^{W^+} & \equiv &
\Delta\sigma_{\vec{p}\,\vec{p}}^{W^+}\, /\,
     \sigma_{pp}^{W^+}\\
& \stackrel{\rm LO}{\simeq} & -\frac{\Delta u(1)\Delta\bar{d}(2)
     +\Delta\bar{d}(1)\Delta u(2)} {u(1)\bar{d}(2)+\bar{d}(1)u(2)}\quad
\xrightarrow{y\text{\raisebox{-0.1cm}{$\stackrel{>}{\sim}$}}+\frac{1}{2}}\quad
    - \frac{\Delta u(1)}{u(1)}\,\,
\frac{\Delta\bar{d}(2)}{\bar{d}(2)}\nonumber\\
\vspace{-0.5cm}
A_{\vec{p}\,\vec{p}}^{W^-} & \equiv &
\Delta\sigma_{\vec{p}\,\vec{p}}^{W^-}\, /\,
     \sigma_{pp}^{W^-}\\
& \stackrel{\rm LO}{\simeq} & -\frac{\Delta d(1)\Delta\bar{u}(2)
     +\Delta\bar{u}(1)\Delta d(2)}{d(1)\bar{u}(2)+\bar{u}(1)d(2)}\quad
\xrightarrow{y\text{\raisebox{-0.1cm}{$\stackrel{>}{\sim}$}}+\frac{1}{2}}\quad
    - \frac{\Delta d(1)}{d(1)}\,\,
\frac{\Delta\bar{u}(2)}{\bar{u}(2)}\nonumber
\end{eqnarray}}
\vspace{-0.5cm}

\noindent with obvious generalizations to $A_{\vec{p}\,d}^{W^{\pm}}$ and
$A_{\vec{p}\,\vec{d}}^{W^{\pm}}$, and where
\begin{equation}
x_{1,2}^0 = (M_W^2/S)^{\frac{1}{2}}\, e^{\pm y}\, .
\end{equation}
In addition we also study the following ratios of singly and doubly
polarized cross sections
{\setlength{\jot}{0.4cm}
\begin{eqnarray}
\frac{\Delta\sigma_{\vec{p}\,p}^{W^+}}
       {\Delta\sigma_{\vec{p}\,p}^{W^-}}
\stackrel{\rm LO}{\simeq} \,\,
\frac{-\Delta u(1)\bar{d}(2) +\Delta\bar{d}(1)u(2)}
  {-\Delta d(1)\bar{u}(2)+\Delta\bar{u}(1)d(2)}
\quad\quad\quad\quad\quad\quad\quad
&
\xrightarrow{y\text{\raisebox{-0.1cm}{$\stackrel{<}{\sim}$}}-\frac{1}{2}}&
\frac{\Delta\bar{d}(1)}{\Delta\bar{u}(1)}\,\,\,
  \frac{u(2)}{d(2)}\\
\frac{\Delta\sigma_{\vec{p}\,d}^{W^+}}
       {\Delta\sigma_{\vec{p}\,d}^{W^-}}
\stackrel{\rm LO}{\simeq} \,\,
\frac{-\Delta u(1)[\bar{u}(2) +\bar{d}(2)] +\Delta\bar{d}(1)
    [u(2)+d(2)]}
  {-\Delta d(1)[\bar{u}(2)+\bar{d}(2)] + \Delta\bar{u}(1)
   [u(2)+d(2)]}
&
\xrightarrow{y\text{\raisebox{-0.1cm}{$\stackrel{<}{\sim}$}}-\frac{1}{2}}&
\frac{\Delta\bar{d}(1)}{\Delta\bar{u}(1)}
\end{eqnarray}}
{\setlength{\jot}{0.4cm}
\begin{eqnarray}
\frac{\Delta\sigma_{\vec{p}\,\vec{p}}^{W^+}}
       {\Delta\sigma_{\vec{p}\,\vec{p}}^{W^-}}
\stackrel{\rm LO}{\simeq} \,\,
\frac{\Delta u(1)\Delta\bar{d}(2) +\Delta\bar{d}(1)\Delta u(2)}
  {\Delta d(1)\Delta\bar{u}(2)+\Delta\bar{u}(1)\Delta d(2)}
\quad\quad\quad\quad\quad\quad\quad\quad\quad
& \xrightarrow{y\text{\raisebox{-0.1cm}{$\stackrel{>}{\sim}$}}+\frac{1}{2}}&
\frac{\Delta u(1)}{\Delta d(1)}\,\,
  \frac{\Delta\bar{d}(2)}{\Delta\bar{u}(2)}\\
\frac{\Delta\sigma_{\vec{p}\,\vec{d}}^{W^+}}
       {\Delta\sigma_{\vec{p}\,\vec{d}}^{W^-}}
\stackrel{\rm LO}{\simeq} \,\,
\frac{\Delta u(1)[\Delta\bar{u}(2) +\Delta\bar{d}(2)] +\Delta\bar{d}(1)
    [\Delta u(2)+d(2)]}
  {\Delta d(1)[\Delta\bar{u}(2)+\Delta\bar{d}(2)] + \Delta\bar{u}(1)
   [\Delta u(2)+\Delta d(2)]}
& \xrightarrow{y\text{\raisebox{-0.1cm}{$\stackrel{<}{\sim}$}}-\frac{1}{2}}&
\frac{\Delta\bar{d}(1)}{\Delta\bar{u}(1)}\,.\quad\quad\quad\quad\quad\quad
\end{eqnarray}}
\vspace{-0.6cm}

\noindent The singly polarized asymmetries in (3.8) and (3.9) are,
for negative
values of $y$, dominated by the polarized antiquark distributions.
(Note that $y=-1$ corresponds to $x_2^0\simeq 0.06$.) The limiting
values $\Delta q/q$ and $\Delta\bar{q}/\bar{q}$ of
$A_{\vec{p}\,p}^{W^{\pm}}$ at the scale $\mu_F^2=M_W^2$ are shown
in fig.\ 5 where they are compared with the flavor--unbroken (AAC)
scenario as well.  It is conceivable that such differences can be
delineated
by future RHIC experiments taken into account their expected
statistical
accuracy \cite{ref22}.

In fig.\ 6 we present the expected double spin asymmetries $A_{\vec{p}\,
\vec{p}}^{W^{\pm}}$ as defined in (3.10) and (3.11) in the broken and
unbroken (AAC) sea scenarios, as specified in fig.\ 1, in LO as well
as in NLO of perturbative QCD.  Here, at large scales $\mu_F\sim M_W$,
the perturbative stability with respect to sizeable variations of the
factorization scale $\mu_F$ is even more pronounced than for the
Drell--Yan ($\gamma^*$) asymmetries in fig.\ 3.  The expected statistical
errors which are reduced as compared to their size in the Drell--Yan
($\gamma^*$) production asymmetries are shown in fig.\ 3.  It is seen that
in any case, LO or NLO estimates, a distinction between both scenarios
is possible due to the sizeable reduction in the statistical errors
involved which is mainly due to the increased luminosity at $\sqrt{S}=
500$ GeV as compared to $\sqrt{S}=50$ GeV, relevant for Drell-Yan
($\gamma^*$) production as discussed in Section 3.1. We recall here
the envisaged integrated luminosities at RHIC, i.e.\ ${\cal{L}}(\sqrt{S})
=(\sqrt{S}/500$ GeV) 800 $pb^{-1}$ and eq.\ (3.5).  In fig.\ 7 we
present the corresponding single asymmetries $A_{\vec{p}\,p}^{W^{\pm}}$
of eqs.\ (3.8) and (3.9) which turn out be less sensitive to the
flavor--broken sea densities than the double spin asymmetries at $y>0$
in fig.\ 6.  An interesting feature demonstrated here is the quality
of the `crude' approximations in eqs.\ (3.8) and (3.9) which are well
satisfied at
$|y|$ \raisebox{-0.1cm}{$\stackrel{>}{\sim}$} $\frac{1}{2}$
in the LO calculations where they are relevant. The ratios of singly
and doubly polarized cross sections in eqs.\ (3.14) and (3.15), together
with their limiting LO  `crude' approximations, are shown in fig.\ 8.
Again, on account of the rather small expected statistical errors,
it is likely that future measurements can discriminate between the
flavor--broken (solid curves) and unbroken (dashed curves) light--sea
scenarios.

Finally we present the following ratio of combinations of polarized
and unpolarized cross sections \cite{ref26}, which turn out to be less
sensitive to absolute normalization uncertainties,
\vspace{-0.3cm}
{\setlength{\jot}{0.5cm}
\begin{eqnarray}
a_{\vec{p}\,N}^W  & \equiv & \,\,
\frac{\Delta\sigma_{\vec{p}\,p}^{W^+} + \Delta\sigma_{\vec{p}\,p}^{W^-}
-\left(
\Delta\sigma_{\vec{p}\,n}^{W^+}+\Delta\sigma_{\vec{p}\,n}^{W^-}\right)}
{\sigma_{pp}^{W^+} +\sigma_{pp}^{W^-} + \sigma_{pn}^{W^+}
+\sigma_{pn}^{W^-}}
\\
& \stackrel{\rm LO}{\simeq} &
\frac{ [\Delta u(1)-\Delta d(1)]\, [\bar{u}(2)-\bar{d}(2)] -
    [\Delta\bar{u}(1)-\Delta\bar{d}(1)]\, [u(2)-d(2)]}
 {[u(1)+d(1)]\, [\bar{u}(2)+\bar{d}(2)] + [\bar{u}(1)+\bar{d}(1)]\,
   [u(2)+d(2)]}\nonumber
\end{eqnarray}
\vspace{-0.9cm}

\noindent and its corresponding doubly polarized counterpart
{\setlength{\jot}{0.5cm}
\begin{eqnarray}
a_{\vec{p}\,\vec{N}}^W  & \equiv & \,\,
\frac{\Delta\sigma_{\vec{p}\,\vec{p}}^{W^+}
    + \Delta\sigma_{\vec{p}\,\vec{p}}^{W^-}
-\left( \Delta\sigma_{\vec{p}\,\vec{n}}^{W^+}
     +\Delta\sigma_{\vec{p}\,\vec{n}}^{W^-}\right)}
{\sigma_{pp}^{W^+} +\sigma_{pp}^{W^-}
+ \sigma_{pn}^{W^+} +\sigma_{pn}^{W^-}}
\\
& \stackrel{\rm LO}{\simeq} &
\frac{ [\Delta u(1)-\Delta d(1)]\, [\Delta\bar{u}(2)-\Delta\bar{d}(2)] -
    [\Delta\bar{u}(1)-\Delta\bar{d}(1)]\, [\Delta u(2)-\Delta d(2)]}
 {[u(1)+d(1)]\, [\bar{u}(2)+\bar{d}(2)] + [\bar{u}(1)+\bar{d}(1)]\,
   [u(2)+d(2)]}\,\, .\nonumber
\end{eqnarray}
\vspace{-0.9cm}

\noindent Both asymmetries have {\em{no}} polarized gluon
contribution
proportional to $\Delta g$ in NLO, which cancels in the various
differences $\Delta\sigma_{\vec{p}\,p}-\Delta\sigma_{\vec{p}\, n}$
and $\Delta\sigma_{\vec{p}\,\vec{p}}\,-\Delta\sigma_{\vec{p}\,\vec{n}}$.
Furthermore, the {\em{double}} helicity asymmetry in (3.18)
{\em{vanishes}} for $\Delta\bar{u}=\Delta\bar{d}$ in LO
{\em{as well as}} in NLO and is thus an interesting combination
to observe the effects of a flavor--broken sea due to $\Delta\bar{u}
\neq\Delta\bar{d}$.  This is explicitly demonstrated in fig.\ 9 where
the sizeably different expectations of the flavor--broken sea scenario
should be easily discernible experimentally.

Finally, for $Z^0$ production we have found that the following two
double and single spin asymmetries are best suited for the
investigation of the flavor structure of the sea:
\begin{eqnarray}
A_{\vec{p}\,\vec{p}}^{Z^0} & \equiv &
  \Delta\sigma_{\vec{p}\,\vec{p}}^{Z^0}\, /\,
   \sigma_{pp}^{Z^0}
\\
& \stackrel{\rm LO}{\simeq} &
-\,\frac{\alpha_u [\Delta u(1)\Delta\bar{u}(2)+\Delta\bar{u}(1)\Delta u(2)]
    +\alpha_d [\Delta d(1)\Delta\bar{d}(2)+\Delta\bar{d}(1)\Delta d(2)]}
       {\alpha_u [u(1)\bar{u}(2)+\bar{u}(1)u(2)] +
        \alpha_d [d(1)\bar{d}(2)+\bar{d}(1)d(2)]}\nonumber
\end{eqnarray}
\begin{eqnarray}
A_{\vec{p},\,p-n}^{Z^0} & \equiv &
  \frac{\Delta\sigma_{\vec{p}\,p}^{Z^0} -
   \Delta\sigma_{\vec{p}\,n}^{Z^0}}
   {\sigma_{pp}^{Z^0} +\sigma_{pn}^{Z^0}}
\\
& \stackrel{\rm LO}{\simeq} &
  \frac{ -[\beta_u\Delta u(1)-\beta_d\Delta d(1)] \,
   [\bar{u}(2) -\bar{d}(2)] \,-\, [\beta_u\Delta\bar{u}(1)-\beta_d
     \Delta\bar{d}(1)]\, [u(2)-d(2)]}
       {[\alpha_u u(1)+\alpha_d d(1)] \,[\bar{u}(2)+\bar{d}(2)] +
         [\alpha_u\bar{u}(1)+\alpha_d\bar{d}(1)]\,
           [u(2)+d(2)]}\nonumber
\end{eqnarray}
\vspace{-0.8cm}

\noindent where $\alpha_q\equiv v_q^2+a_q^2$ and $\beta_q\equiv2v_q a_q$
as explained and given in the Appendix.  These asymmetries are
depicted, together with their expected statistical errors, in
fig.\ 10.  It should be emphasized that these $Z^0$--production
asymmetries are up to an order of magnitude larger in the
flavor--broken ($\Delta\bar{u}\neq\Delta\bar{d}$) scenario than
for flavor--symmetric sea densities ($\Delta\bar{u}=\Delta\bar{d}$)
where they become almost unmeasurably small.

\section{Summary and Conclusions}

The possibility to determine the flavor structure of the polarized
antiquark (sea) distributions of the nucleon via vector boson
($\gamma^*,\, W^{\pm},\, Z^0$) production at high energy polarized
hadron--hadron ($\vec{p}\,\vec{p},\, \vec{p}\,\vec{n}\,(\vec{d}\,)$)
colliders was investigated.  The perturbative stability of the
expected asymmetries for two representative models for the flavor
structure of the sea distributions was studied and has shown that
the predicted distinctive signatures for both flavor--symmetric and
flavor--asymmetric models remain essentially unchanged in LO and
NLO of perturbative QCD.  This demonstrates that these characteristic
and distinctive features are genuine signatures of the models under
consideration for the flavor structure of the polarized sea.

In particular the polarized Drell--Yan ($\gamma^*$) dilepton
production asymmetry $A_{\vec{p}-\vec{n}}^{\gamma^*}$ in (3.2), as
obtained from $\vec{p}\,\vec{p}$ and $\vec{p}\,\vec{n}$ collisions,
or the ratio $R_{pd}$ of polarized $\vec{p}\,\vec{p}$ and
$\vec{p}\,\vec{n}$ production cross sections in (3.3) provide us
with characteristic and direct signatures for a flavor--broken
polarized sea, $\Delta\bar{u}\neq\Delta\bar{d}$, as illustrated in
fig.\ 3.  At much larger factorization scales $\mu_F^2\sim M_{W,\,Z}^2$,
the double spin asymmetries $A_{\vec{p}\,\vec{p}}^{W^{\pm}}$ for
$W^{\pm}$ production in (3.10) and (3.11), or combinations of doubly
polarized $\vec{p}\,\vec{p}$ and $\vec{p}\,\vec{n}$
$W^{\pm}$--production cross sections in (3.18) constitute similar
clean and distinctive observables for studying the flavor structure
of the polarized light sea, as shown in figs.\ 6 and 9.  Somewhat
less sensitive signatures for a flavor--broken sea are provided by
the single spin asymmetries $A_{\vec{p}\,p}^{W^{\pm}}$ in (3.8) and
(3.9), although they may give access directly to $\Delta\bar{d}(x,
M_W^2)$ and $\Delta\bar{u}(x,M_W^2)
$ in specific kinematic regions
($y$ \raisebox{-0.1cm}{$\stackrel{<}{\sim}$} $-0.5$).
Such direct signatures for $\Delta\bar{d}$ and $\Delta\bar{u}$ could
also be obtained from studying ratios of $W^+$ and $W^-$ cross
sections of singly and doubly polarized $pd$ collisions, cf.\ (3.14)
and (3.16).  Finally, the double spin asymmetry
$A_{\vec{p}\,\vec{p}}^{Z^0}$ production at RHIC is an equally
useful observable for delineating the flavor structure of the polarized
sea, since it is expected to be about an order of magnitude larger
for a flavor--broken ($\Delta \bar{u}\neq\Delta\bar{d}$) than for a
flavor--symmetric ($\Delta \bar{u}=\Delta\bar{d}$) polarized sea
scenario as shown in fig.\ 10.

The resolution power of the asymmetries studied depends of course
on the expected statistical errors which were estimated for the
envisaged beam polarizations and luminosities at RHIC.  They point
towards the superiority of vector boson ($W^{\pm},\, Z^0$) production
(figs.\ 6 -- 10) over the common Drell-Yan $\gamma^*$ production
(fig.\ 3) as a tool for studying the flavor structure of the
polarized sea distributions.  This derives mainly from the increased
luminosity at the corresponding higher energies involved in vector
boson production, i.e.\ $\sqrt{S}\simeq 500$ GeV, as compared to
$\sqrt{S}\simeq 50$ GeV relevant for $\gamma^*$ (dilepton) production
as discussed in Section 3.1.

\section*{Appendix}
\setcounter{equation}{0}
\renewcommand{\theequation}{A.\arabic{equation}}
\def\xa{x_{1}}
\def\xb{x_{2}}
\def\xaa{{x_{1}^{0}}}
\def\xbb{x_{2}^{0}}
\def\d{{\rm d}}
\def\Li{{\rm Li}}
\allowdisplaybreaks

Here we summarize all those unpolarized and polarized  cross sections for
(Drell-Yan) vector boson ($\gamma^{\ast}$,$W^{\pm}$,$Z^{0}$) production in
LO and NLO($\overline{\text{MS}}$) needed for calculating the various
spin-asymmetries suggested and studied in this paper.

In terms of cross sections of definite positive and negative hadron
helicities ($\pm$), an \emph{unpolarized} cross section is generally defined
by $\sigma=\frac{1}{4}(\sigma_{++}+\sigma_{+-}+\sigma_{-+}+\sigma_{--})$.
The relevant differential unpolarized Drell-Yan cross section for $h_1 h_2
\rightarrow \gamma^{\ast} X \rightarrow l^+ l^- X$ can be written as
\begin{eqnarray}
M^2 \frac{\d \sigma_{h_1 h_2}^{\gamma^{\ast}}(x_F,M^2,\mu_F^2)}{\d M^2 \ \d
x_F} & = &  N^{\gamma^{\ast}} \sum_{q=u,d,s} e^{2}_{q}
\int_{x_1^0}^1 \d x_1 \int_{x_2^0}^1 \d x_2 \nonumber \\
& &  \hspace{-1.6cm}
\times \Bigg\{\left[
D_{q\bar q}^{(0)} (x_1,x_2,\xaa,\xbb)  +\frac{\alpha_s}{2\pi}
D_{q\bar{q}}^{(1)}
\left(x_1,x_2,\xaa,\xbb,\frac{M^2}{\mu_F^2}\right)\right]\nonumber \\
& & \hspace{-0.4cm} \times
\Big\{ q(x_1,\mu_F^2) \bar{q}(x_2,\mu_F^2) +
\bar{q}(x_1,\mu_F^2) q(x_2,\mu_F^2) \Big\} \nonumber \\
& & \hspace{-1.2cm} +  \frac{\alpha_s}{2\pi}
D_{gq}^{(1)} \left(x_1,x_2,\xaa,\xbb,
\frac{M^2}{\mu_F^2}\right) g(x_1,\mu_F^2) \left\{
q(x_2,\mu_F^2) +
\bar{q} (x_2,\mu_F^2) \right\} \nonumber \\
& & \hspace{-1.2cm} + \frac{\alpha_s}{2\pi}
D_{qg}^{(1)} \left(x_1,x_2,\xaa,\xbb,
\frac{M^2}{\mu_F^2}\right) \left\{
q(x_1,\mu_F^2) +
\bar{q} (x_1,\mu_F^2) \right\} g(x_2,\mu_F^2) \Bigg\} \nonumber \\
\end{eqnarray}
with $N^{\gamma^{\ast}}=4 \pi \alpha^2 / 9 S$, $\alpha_s=\alpha_s(\mu_F^2)$,
$D_{q\bar q}^{(0)} (x_1,x_2,\xaa,\xbb)=\delta (x_1-x_1^0)\delta (x_2-x_2^0)/
(x_1^0+x_2^0)$ and according to the NLO($\overline{\text{MS}}$) results of
[28,23]
\begin{eqnarray}
D_{q\bar{q}}^{(1)}\left(x_1,x_2,\xaa,\xbb,\frac{M^2}{\mu_F^2}\right)
& = & C_F \Bigg\{ \frac{\delta (x_1-x_1^0) \,
\delta (x_2-x_2^0)}{x_1^0+x_2^0} \bigg[ \frac{\pi^2}{3} - 8 + 2 \Li_2 (\xaa
)
+ 2 \Li_2 ( \xbb )
\nonumber \\
& & \hspace{0.6cm}
+\ln^2 (1-\xaa) + \ln^2 (1-\xbb) + 2 \ln \frac{\xaa}{1-\xaa} \ln
\frac{\xbb}{1-\xbb} \bigg]  \nonumber \\
& & + \Bigg(\frac{\delta (\xa-\xaa)}{x_1^0+x_2^0} \bigg[ \frac{1}{\xb}
- \frac{\xbb}{\xb^2} - \frac{\xbb\,^2+\xb^2}{\xb^2(\xb-\xbb)} \ln
\frac{\xbb}{\xb} \nonumber \\
& & \hspace{0.6cm}
+ \frac{\xbb\,^2+\xb^2}{\xb^2}
\left(\frac{\ln (1-\xbb/\xb)}{\xb-\xbb}\right)_{+}
+ \frac{\xbb\,^2+\xb^2}{\xb^2} \frac{1}{\left(\xb-\xbb\right)_{+}}
\nonumber\\ && \hspace{0.6cm} \ln \frac
{(\xaa+\xbb)(1-\xaa)}{\xaa (\xaa+\xb)}
\bigg] + (1 \leftrightarrow 2 ) \Bigg) \nonumber \\
& & + \frac{
G^A(\xa,\xb,\xaa,\xbb)}{\left[(\xa-\xaa)(\xb-\xbb)\right]_{+}} +
H^A (\xa,\xb,\xaa,\xbb)
\nonumber \\
&&
+ \ln \frac{M^2}{\mu_F^2} \Bigg\{  \frac{\delta (x_1-x_1^0)\,
\delta (x_2-x_2^0)}{x_1^0+x_2^0} \bigg[ 3 + 2 \ln \frac{1-\xaa}{\xaa} + 2
\ln
\frac{1-\xbb}{\xbb}
\bigg] \nonumber\\
&& \hspace{0.6cm}
+ \bigg( \frac{\delta (\xa-\xaa)}{x_1^0+x_2^0}
 \frac{\xbb\,^2+\xb^2}{\xb^2}\frac{1}{\left(\xb-\xbb\right)_{+}} + (1
\leftrightarrow 2 ) \bigg) \Bigg\} \Bigg\} \\
D_{qg}^{(1)}\left(x_1,x_2,\xaa,\xbb,\frac{M^2}{\mu_F^2}\right)
& = &
T_F\Bigg\{\frac{\delta (\xb-\xbb) }{(\xaa+\xbb)\xa^3} \Bigg[ (\xaa^2
+(\xa-\xaa)^2)
\ln\frac{(\xaa+\xbb)(1-\xbb)(\xa-\xaa)}{\xaa\xbb(\xa+\xbb)}
 \nonumber \\
& & \hspace{0.6cm}
+ 2 \xaa (\xa-\xaa)\Bigg]
+ \frac{G^C(\xa,\xb,\xaa,\xbb)}{(\xb-\xbb)_{+}} +
H^C(\xa,\xb,\xaa,\xbb)
\nonumber \\ & &
+  \ln \frac{M^2}{\mu_F^2} \Bigg\{ \frac{\delta (\xb-\xbb)
}{(\xaa+\xbb)\xa^3}
 (\xaa^2 +(\xa-\xaa)^2) \Bigg\}\Bigg\} \nonumber\\
\end{eqnarray}
where $C_F=4/3$, $T_F=1/2$, $\Li_2 (x) = - \int^{x}_{0} \d t \ln(1-t)/t$ and
\begin{eqnarray}
G^A(\xa,\xb,\xaa,\xbb) & = & \frac{(\xa+\xb)(\xaa\,^2\xbb\,^2+\xa^2\xb^2)}
{\xa^2\xb^2(\xaa+\xb)(\xa+\xbb)} \nonumber \\
H^A (\xa,\xb,\xaa,\xbb) & = &
-\frac{2}{\xa\xb(\xa+\xb)} \nonumber \\
G^C(\xa,\xb,\xaa,\xbb) & = &
\frac{2\xaa\xbb-\xa\xb}{\xa^2\xb(\xaa+\xb)}\nonumber \\
H^C (\xa,\xb,\xaa,\xbb) & = &\frac{\xa(\xaa+\xb)(\xb-\xbb)+2\xaa\xbb
(\xa+\xb)}{\xa^2\xb^2(\xa+\xb)^2}\ .
\end{eqnarray}

Furthermore
\begin{equation}
D_{gq}^{(1)}\left(x_1,x_2,\xaa,\xbb,\frac{M^2}{\mu_F^2}\right) =
D_{qg}^{(1)}\left(x_2,x_1,\xbb,\xaa,\frac{M^2}{\mu_F^2}\right)
\end{equation}
with the factorization scale $\mu_F$, usually assumed to be given by $M$,
and $x^0_{1,2}$ given by (3.4), with the constraints $\tau \le x^0_{1,2} \le
1$ and $-1+\tau \le x_F \le 1-\tau$. Alternatively one may define
$x^0_{1,2}=\sqrt{\tau} e^{\pm y}$ with the lepton pair rapidity $y$ in the
hadron-hadron c.m. system being constrained by $\ln\sqrt{\tau} \le y \le
-\ln \sqrt{\tau}$. The unpolarized cross sections for vector boson
($VB=W^{\pm},Z^0$) production can be written as (for simplicity we use the
same symbols for the coefficient functions as in (A.1) although they
obviously differ for different processes)
\begin{eqnarray}
\frac{\d \sigma_{h_1 h_2}^{VB}(y,\mu_F^2)}{\d y} & = &  N^{VB} \sum_{q,q'}
c^{VB}_{qq'}
\int_{x_1^0}^1 \d x_1 \int_{x_2^0}^1 \d x_2 \nonumber \\
& &  \hspace{-1.6cm}
\times \Bigg\{\left[
D_{q\bar q}^{(0)} (x_1,x_2,\xaa,\xbb)  +\frac{\alpha_s}{2\pi}
D_{q\bar{q}}^{(1)}
\left(x_1,x_2,\xaa,\xbb,\frac{M_{VB}^2}{\mu_F^2}\right)\right]\nonumber \\
& & \hspace{-0.4cm} \times
\Big\{ q(x_1,\mu_F^2) \bar{q}'(x_2,\mu_F^2) +
\bar{q}(x_1,\mu_F^2) q'(x_2,\mu_F^2) \Big\} \nonumber \\
& & \hspace{-1.2cm} +  \frac{\alpha_s}{2\pi}
D_{gq}^{(1)} \left(x_1,x_2,\xaa,\xbb,
\frac{M_{VB}^2}{\mu_F^2}\right) g(x_1,\mu_F^2) \left\{
q'(x_2,\mu_F^2) +
\bar{q}' (x_2,\mu_F^2) \right\} \nonumber \\
& & \hspace{-1.2cm} + \frac{\alpha_s}{2\pi}
D_{qg}^{(1)} \left(x_1,x_2,\xaa,\xbb,
\frac{M_{VB}^2}{\mu_F^2}\right) \left\{
q(x_1,\mu_F^2) +
\bar{q}(x_1,\mu_F^2) \right\} g(x_2,\mu_F^2) \Bigg\} \nonumber\\
\end{eqnarray}
with $N^{VB}= \sqrt{2} \pi G_F M^2_{VB}/ 3 S$, using $M_W=80.42\ GeV$ and
$M_Z=91.19\ GeV$, and $c^{W^{\pm}}_{qq'}=|V_{qq'}|^2$ with the relevant CKM
matrix elements $V_{ud} \approx 0.97$ and $V_{us} \approx 0.22$, and
$c^{Z^0}_{qq'}=\delta_{qq'}\ (v^2_q+a^2_q)$ with $v^2_u+a^2_u \approx 0.29$
and $v^2_d+a^2_d \approx 0.37$. Furthermore we have now $D_{q\bar q}^{(0)}
(x_1,x_2,\xaa,\xbb)=\delta (x_1-x_1^0)\delta (x_2-x_2^0)$ and the
NLO($\overline{\text{MS}}$) coefficients $D_{q\bar q}^{(1)}$ and
$D_{qg}^{(1)}$ are given by eqs. (15) and (16) of ref. [29], with
$D_{gq}^{(1)}$ being again given by (A.5).

The \emph{doubly} longitudinally \emph{polarized} cross sections are
generally defined via $\Delta \sigma_{\vec{h}_1 \vec{h}_2}=\frac{1}{4}
(\sigma_{++}-\sigma_{+-}-\sigma_{-+}+\sigma_{--})$.  The relevant
differential polarized Drell-Yan cross section for $\vec{h}_1 \vec{h}_2
\rightarrow \gamma^{\ast} X \rightarrow l^+ l^- X$ is given by (1,2 denote
the arguments $x_{1,2}$,$\mu_F$ and $\ldots$ denotes the variables of the
coefficient functions in (A.2-A.6) )
\begin{eqnarray}
M^2 \frac{\d \Delta \sigma_{\vec{h}_1
\vec{h}_2}^{\gamma^{\ast}}(x_F,M^2,\mu_F^2)}{\d M^2 \ \d x_F} & = & -
N^{\gamma^{\ast}} \sum_{q=u,d,s} e^2_{q}
\int_{x_1^0}^1 \d x_1 \int_{x_2^0}^1 \d x_2 \nonumber \\
& &  \hspace{-1.6cm}
\times \Bigg\{\left[
D_{q\bar q}^{(0)} (\ldots)  +\frac{\alpha_s}{2\pi}
D_{q\bar{q}}^{(1)}
\left(\ldots\right)\right]\nonumber \\
& & \hspace{-0.4cm} \times
\Big\{ \Delta q(1) \Delta \bar{q}(2)   +
\Delta \bar{q}(1) \Delta q(2) \Big\} \nonumber \\
& & \hspace{-1.2cm} + \frac{\alpha_s}{2\pi}
\Delta D_{gq}^{(1)} \left(\ldots \right) \Delta g(1) \left\{
\Delta q(2) +
\Delta \bar{q}(2) \right\} \nonumber \\
& & \hspace{-1.2cm} +  \frac{\alpha_s}{2\pi}
\Delta D_{qg}^{(1)} \left(\ldots \right) \left\{
\Delta q(1) +
\Delta \bar{q}(1) \right\} \Delta
g(2)  \Bigg\} \nonumber \\
\end{eqnarray}
where $D_{q\bar q}^{(0,1)}$ are as in (A.1) with $D_{q\bar q}^{(1)}$ given
in (A.2) and [23]
\begin{eqnarray}
\Delta D_{qg}^{(1)}\left(x_1,x_2,\xaa,\xbb,\frac{M^2}{\mu_F^2}\right)
& = &
T_F\Bigg\{\frac{\delta (\xb-\xbb) }{(\xaa+\xbb)\xa^2} \Bigg[ (2\xaa -\xa)
\ln\frac{(\xaa+\xbb)(1-\xbb)(\xa-\xaa)}{\xaa\xbb(\xa+\xbb)}
 \nonumber \\
& & \hspace{0.6cm}
+ 2(\xa-\xaa)\Bigg]
+ \frac{\Delta G^C(\xa,\xb,\xaa,\xbb)}{(\xb-\xbb)_{+}} +
H^C(\xa,\xb,\xaa,\xbb)
\nonumber \\ & &
+  \ln \frac{M^2}{\mu_F^2} \Bigg\{ \frac{\delta (\xb-\xbb)
}{(\xaa+\xbb)\xa^2}
 (2\xaa -\xa) \Bigg\}\Bigg\}
\end{eqnarray}
with $H^C(\ldots)$ given in (A.4) and
\begin{eqnarray}
\Delta  G^C(\xa,\xb,\xaa,\xbb) & = &
\frac{(\xaa\xbb)^2+(\xa\xb-\xaa\xbb)^2}{\xa^3\xb^2(\xaa+\xb)}\ .
\end{eqnarray}
Similarly  to (A.5) we have
\begin{equation}
\Delta D_{gq}^{(1)}\left(x_1,x_2,\xaa,\xbb,\frac{M^2}{\mu_F^2}\right) =
\Delta D_{qg}^{(1)}\left(x_2,x_1,\xbb,\xaa,\frac{M^2}{\mu_F^2}\right)\ .
\end{equation}
The doubly longitudinally polarized cross sections for vector boson
production are given by
\begin{eqnarray}
\frac{\d \Delta \sigma^{VB}_{\vec{h}_1 \vec{h}_2} (y,\mu_F^2)}{\d y} & = & -
N^{VB} \sum_{q,q'} c^{VB}_{qq'}
\int_{x_1^0}^1 \d x_1 \int_{x_2^0}^1 \d x_2 \nonumber \\
& &  \hspace{-1.6cm}
\times \Bigg\{\left[
D_{q\bar q}^{(0)} (\ldots)  +\frac{\alpha_s}{2\pi}
D_{q\bar{q}}^{(1)}
\left(\ldots\right)\right]\nonumber \\
& & \hspace{-0.4cm} \times
\Big\{ \Delta q(1) \Delta \bar{q}'(2)   +
\Delta \bar{q}(1) \Delta q'(2) \Big\} \nonumber \\
& & \hspace{-1.2cm} + \frac{\alpha_s}{2\pi}
\Delta D_{gq}^{(1)} \left(\ldots \right) \Delta g(1) \left\{
\Delta q'(2) +
\Delta \bar{q}' (2) \right\} \nonumber \\
& & \hspace{-1.2cm} +  \frac{\alpha_s}{2\pi}
\Delta D_{qg}^{(1)} \left(\ldots \right) \left\{
\Delta q(1) +
\Delta \bar{q} (1) \right\} \Delta
g(2)  \Bigg\} \nonumber \\
\end{eqnarray}
with $N^{VB}$, $c^{VB}_{qq'}$ and $D_{q\bar q}^{(0,1)}$ as in (A.6) and
$\Delta D_{gq}^{(1)}$ is now given by eq. (18) of ref. [29] which relates to
$\Delta D_{qg}^{(1)}$ again via (A.10).

Finally, the \emph{singly} longitudinally \emph{polarized} cross sections
for vector boson ($VB=W^{\pm}$, $Z^0$) production are generally defined via
$\Delta \sigma_{\vec{h}_1 h_2}=\frac{1}{4}
(\sigma_{++}+\sigma_{+-}-\sigma_{-+}-\sigma_{--})$ which are given by [29]
\begin{eqnarray}
\frac{\d \Delta \sigma^{VB}_{\vec{h}_1 h_2}(y,\mu_F^2)}{\d y} & = & N^{VB}
\sum_{q,q'} c^{VB}_{qq'}
\int_{x_1^0}^1 \d x_1 \int_{x_2^0}^1 \d x_2 \nonumber \\
& &  \hspace{-1.6cm}
\times \Bigg\{\left[
D_{q\bar q}^{(0)} (\ldots)  +\frac{\alpha_s}{2\pi}
D_{q\bar{q}}^{(1)}
\left(\ldots \right)\right]\nonumber \\
& & \hspace{-0.4cm} \times
\Big\{ - \Delta q(1) \bar{q}'(2) +
\Delta \bar{q}(1) q'(2) \Big\} \nonumber \\
& & \hspace{-1.2cm} + \frac{\alpha_s}{2\pi}
\Delta D_{gq}^{(1)} \left(\ldots \right) \Delta g(1) \left\{
q'(2) -
\bar{q}' (2) \right\} \nonumber \\
& & \hspace{-1.2cm} +  \frac{\alpha_s}{2\pi}
D_{qg}^{(1)} \left(\ldots \right) \left\{
- \Delta q(1) +
\Delta \bar{q} (1) \right\}
g(2)  \Bigg\}  \nonumber \\
\end{eqnarray}
where all normalizations, couplings and coefficient functions are as in
(A.11) except for $c^{Z^0}_{qq'}$ which is now given by
$c^{Z^0}_{qq'}=\delta_{qq'}\ 2 v_q a_q$, i.e.
$c^{Z^0}_{uu}=\frac{1}{2}-\frac{4}{3} \sin^2 \Theta_W \approx 0.19$ and
$c^{Z^0}_{dd}=c^{Z^0}_{ss}=\frac{1}{2}-\frac{2}{3} \sin^2 \Theta_W \approx
0.35$.

%

\newpage

\noindent{\large{\bf{\underline{Figure Captions}}}}
\begin{itemize}
\item[\bf{Fig.\ 1}.]  Polarized (anti)quark distributions in LO and NLO
      according to eqs.\ (2.2) -- (2.6) in the flavor--broken sea scenario
      (full curves) at various fixed factorization scales $\mu_F^2\equiv
Q^2$.
      For comparison the corresponding `standard' flavor--symmetric sea
      distribution $\Delta\bar{q}$ according to AAC \cite{ref18} is shown
      by the dashed curves.  The remaining $\Delta u$ and $\Delta d$
      densities are similar in both scenarios (cf.\ fig.\ 5).

\item[\bf{Fig.\ 2}.]  The unpolarized Drell-Yan dilepton production cross
      section $d^2\sigma_{pp}^{\gamma^*(M)}/dM\, dx_F$ for $\sqrt{S}=50$
      and 100 GeV in NLO based on the GRV98 \cite{ref21} NLO parton
      distributions.

\item[\bf{Fig.\ 3}.]  The ratio $R_{p+n}$ of polarized $\vec{p}\,\vec{p}$
and
      $\vec{p}\,\vec{n}$ Drell--Yan cross sections in eq.\ (3.3) and the
      asymmetry
      $A_{\vec{p}-\vec{n}}^{\gamma^*(M)}$ in eq.\ (3.2) in LO and NLO based
      on the distributions of the broken and unbroken (AAC) sea scenarios
      as specified in fig.\ 1.  The perturbative stability of the predicted
      asymmetries and their sensitivity to the choice of the factorization
      scale $\mu_F$ is examined in the two lower figures.  Statistical
      errors are calculated according to eqs.\ (3.5) and (3.6).  The `crude
      approx.' curve refers to $[1+\Delta\bar{d}(2)/\Delta\bar{u}(2)]/2$
      in eq.\ (3.3).

\item[\bf{Fig.\ 4}.] The unpolarized differential cross sections
      $d\sigma_{pp}^{W^{\pm},\,Z^0}/dy$ for $\sqrt{S}=500$ GeV (RHIC) in
      NLO based on the GRV98 \cite{ref21} NLO parton distributions.

\item[\bf{Fig.\ 5}.]  The LO (anti)quark asymmetries
     $(\Delta\bar{q}/\bar{q})\,\,\Delta q/q$ at the relevant factorization
     scale $\mu_F^2\equiv Q^2=M_W^2$, dominating the single spin asymmetries
     for $A_{\vec{p}\,p}^{W^{\pm}}$ at $|y|$ \raisebox{-0.1cm}
     {$\stackrel{>}{\sim}$} $\frac{1}{2}$ according to eqs.\ (3.8) and
     (3.9).

\item[\bf{Fig.\ 6}.] The double spin asymmetries
$A_{\vec{p}\,\vec{p}}^{W^+}$
     [eq.\ (3.10)] and $A_{\vec{p}\,\vec{p}}^{W^-}$ [eq.\ (3.11)] at
     $\sqrt{S}=500$ GeV in the broken and unbroken (AAC) sea scenarios.
     The statistical errors are evaluated according to eq.\ (3.5).  The
     perturbative stability of the predicted asymmetries and their
     sensitivity
     to the choice of the factorization scale $\mu_F$ is shown as well.

\item[\bf{Fig.\ 7}.]  The single spin asymmetries $A_{\vec{p}\,p}^{W^+}$
     [eq.\ (3.8)] and $A_{\vec{p}\,p}^{W^-}$ [eq.\ (3.9)] at $\sqrt{S}=500$
     GeV in the broken and unbroken (AAC) sea scenarios, as specified in
     fig.\ 1.  The statistical errors are evaluated according to eq.\ (3.7).
     The errors on the AAC curves are similarly small as in the broken
     scenario.  The quality of the `crude' approximations in eq.\ (3.8),
     $\Delta\bar{d}(1)/\bar{d}(1)$, and eq.\ (3.9), $\Delta\bar{u}(1)/
     \bar{u}(1)$, is examined as well.  The perturbative stability of the
     predicted asymmetries is similar to the one shown in \mbox{fig.\ 6.}

\item[\bf{Fig.\ 8}.] The ratios $\Delta\sigma_{\vec{p}\,\vec{p}}^{W^+}/
     \Delta\sigma_{\vec{p}\,\vec{p}}^{W^-}$
     and $\Delta\sigma_{\vec{p}\,d}^{W^+}/\Delta\sigma_{\vec{p}\,d}^{W^-}$
     for doubly and singly polarized cross sections, respectively, at
     $\sqrt{S}=500$ GeV obtained via eqs.\ (3.15) and (3.14).  The meaning
     of the curves corresponds to that in fig.\ 7.

\item[\bf{Fig. 9}.]  The asymmetries $a_{\vec{p}\,\vec{N}}^W$ and
    $a_{\vec{p}\,N}^W$ at $\sqrt{S}=500$ GeV in eqs.\ (3.18) and (3.17),
    respectively, for the broken and unbroken (AAC) scenario.

\item[\bf{Fig.\ 10}.] The asymmetries $A_{\vec{p}\,\vec{p}}^{Z^0}$ and
    $A_{\vec{p},\, p-n}^{Z^0}$ in LO and NLO at $\sqrt{S}=500$ GeV given
    in eqs.\ (3.19) and (3.20), respectively.  The meaning of the curves
    corresponds to that in fig.\ 9.
\end{itemize}

\newpage
\pagestyle{empty}
\headheight 0cm
\headsep 0cm
\topmargin 0cm
\voffset -15mm
\textheight 27cm
\textwidth 18cm
\oddsidemargin -1cm \evensidemargin -1cm
\topmargin 0.0cm
\begin{figure}[h]

  \centerline{\hbox{
    \psfig{figure=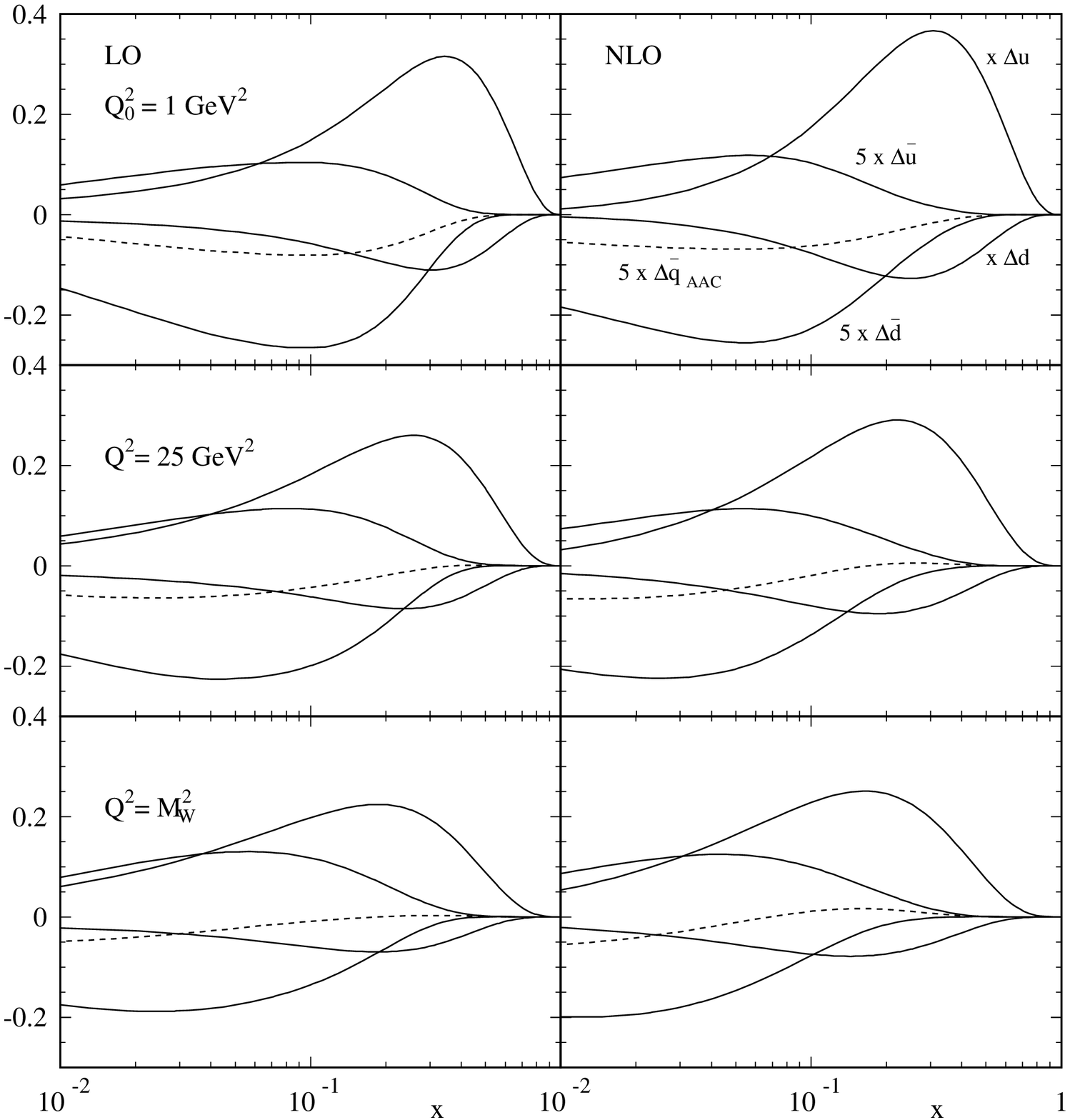,width=12cm,height=14cm,angle=0}
 }}
\vspace*{2cm}
\hspace*{8cm}
{\large \bf Figure 1}
\vspace*{8cm}
\end{figure}

\begin{figure}[H]
  \centerline{\hbox{
    \psfig{figure=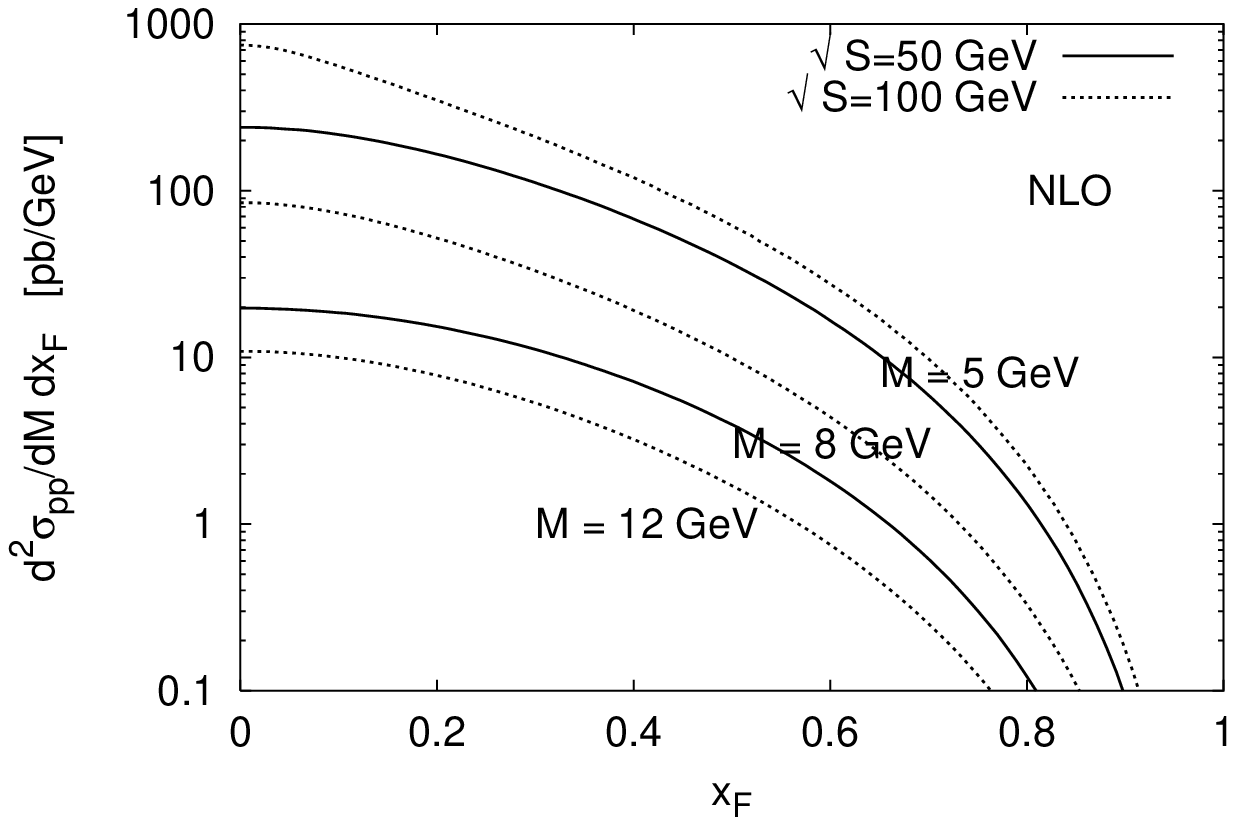,width=12cm,height=8cm,angle=0}
 }}
\hspace*{8cm}
{\large \bf Figure 2}
\end{figure}

\begin{figure}[H]
  \centerline{\hbox{
    \psfig{figure=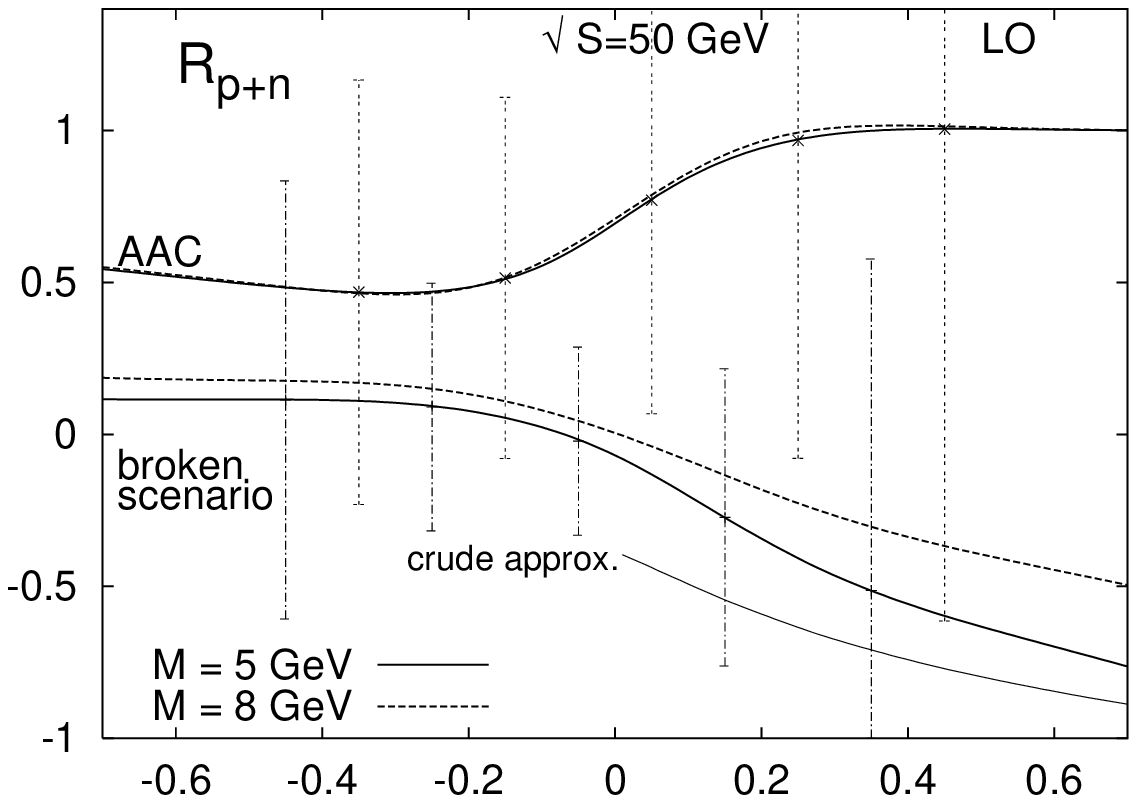,width=9cm,height=6.5cm,angle=0}
     \psfig{figure=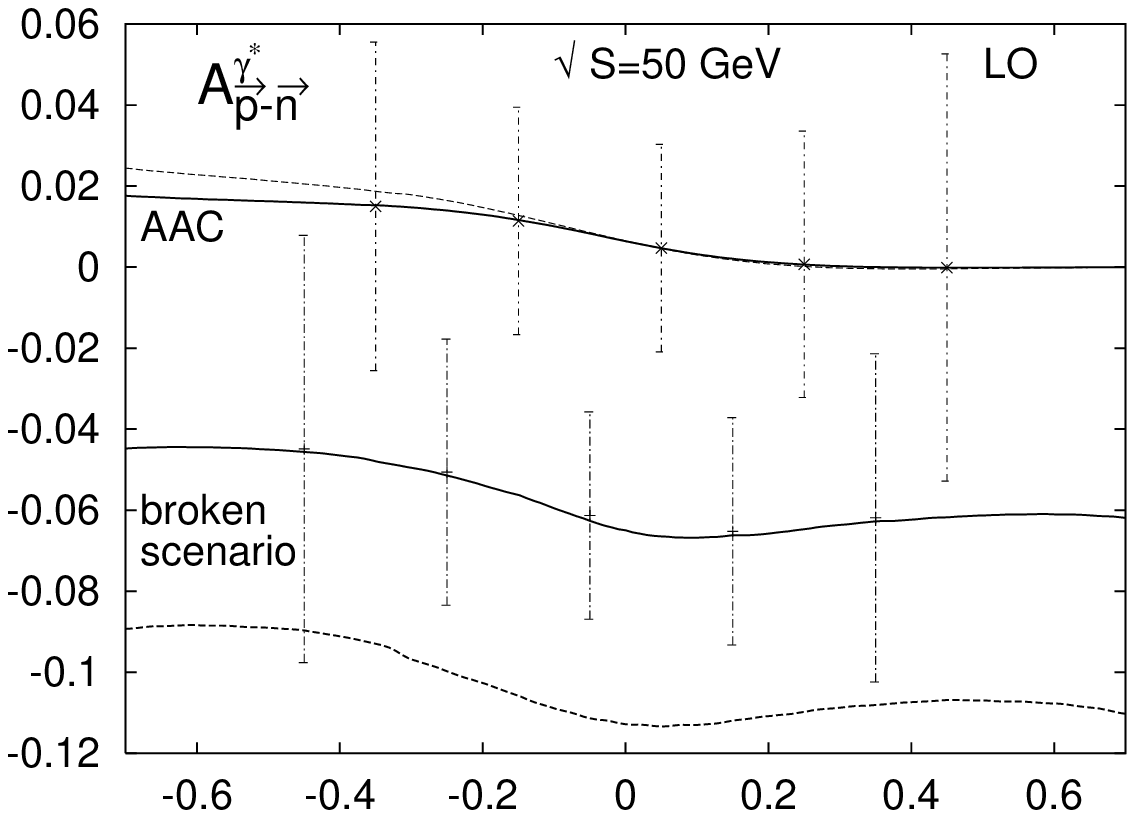,width=9cm,height=6.5cm,angle=0}
 }}
  \centerline{\hbox{
    \psfig{figure=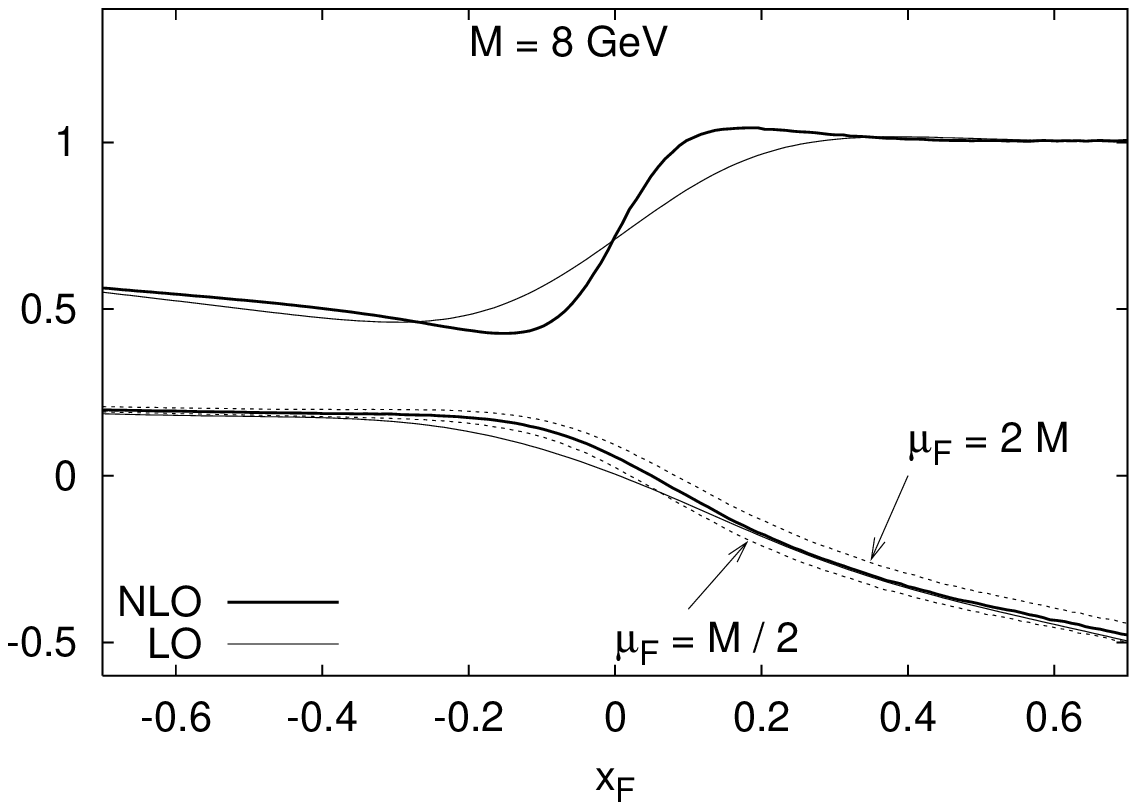,width=9cm,height=6.5cm,angle=0}
     \psfig{figure=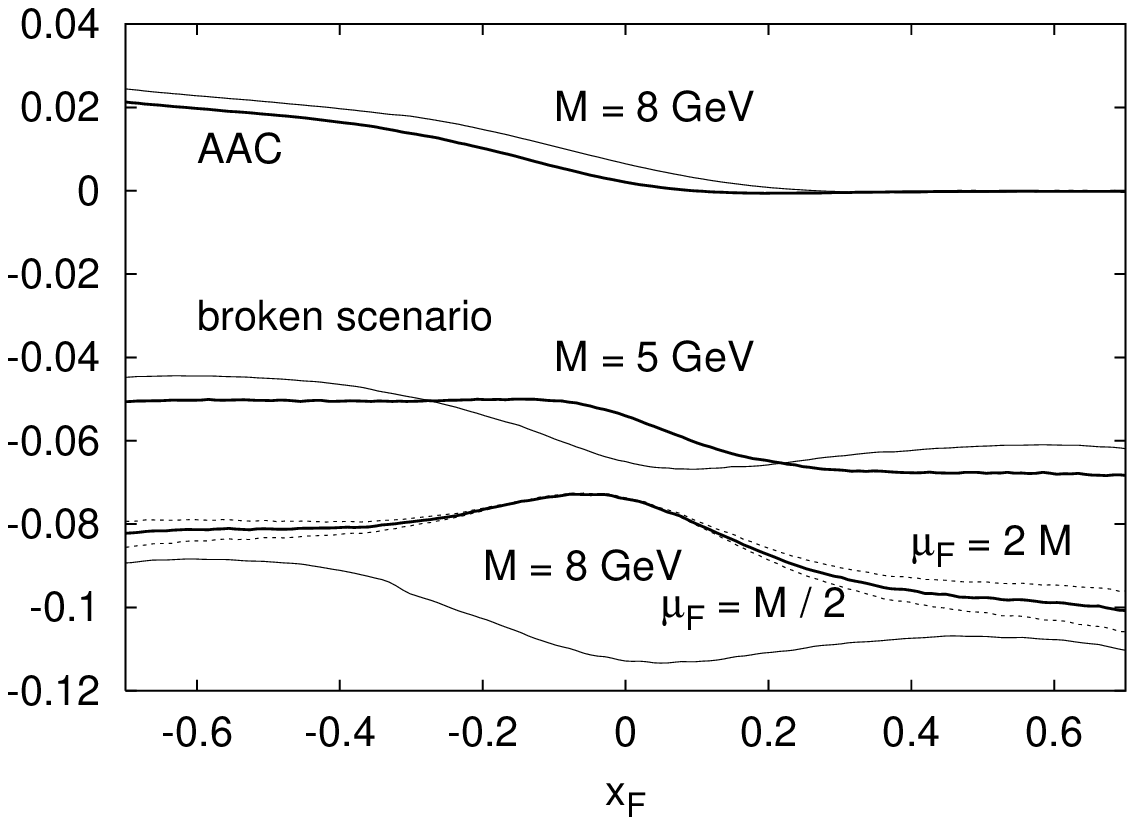,width=9cm,height=6.5cm,angle=0}
 }}
\hspace*{8cm}
{\large \bf Figure 3}
\end{figure}

\vspace*{-2cm}
\begin{figure}[H]
  \centerline{\hbox{
    \psfig{figure=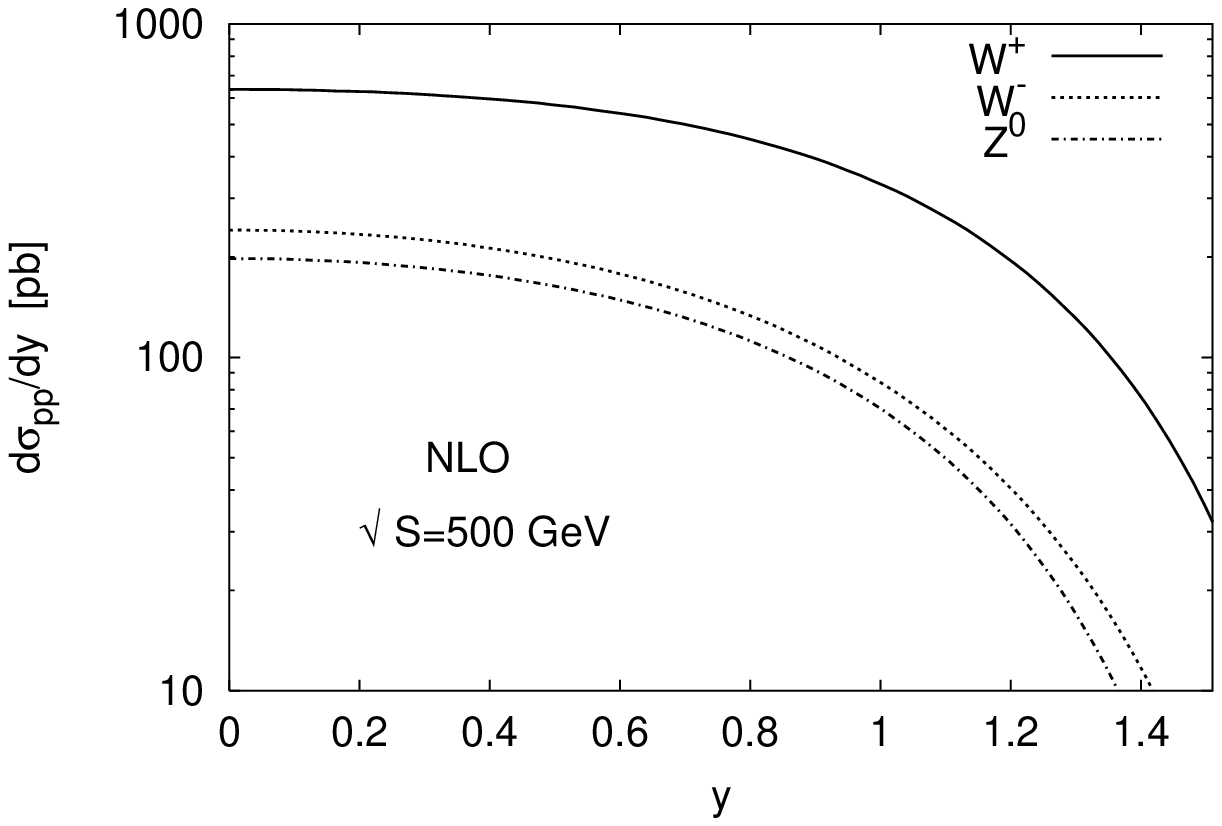,width=13cm,height=9cm,angle=0}
 }}
\hspace*{8cm}
{\large \bf Figure 4}
\end{figure}

\vspace*{1cm}
\begin{figure}[H]
  \centerline{\hbox{
    \psfig{figure=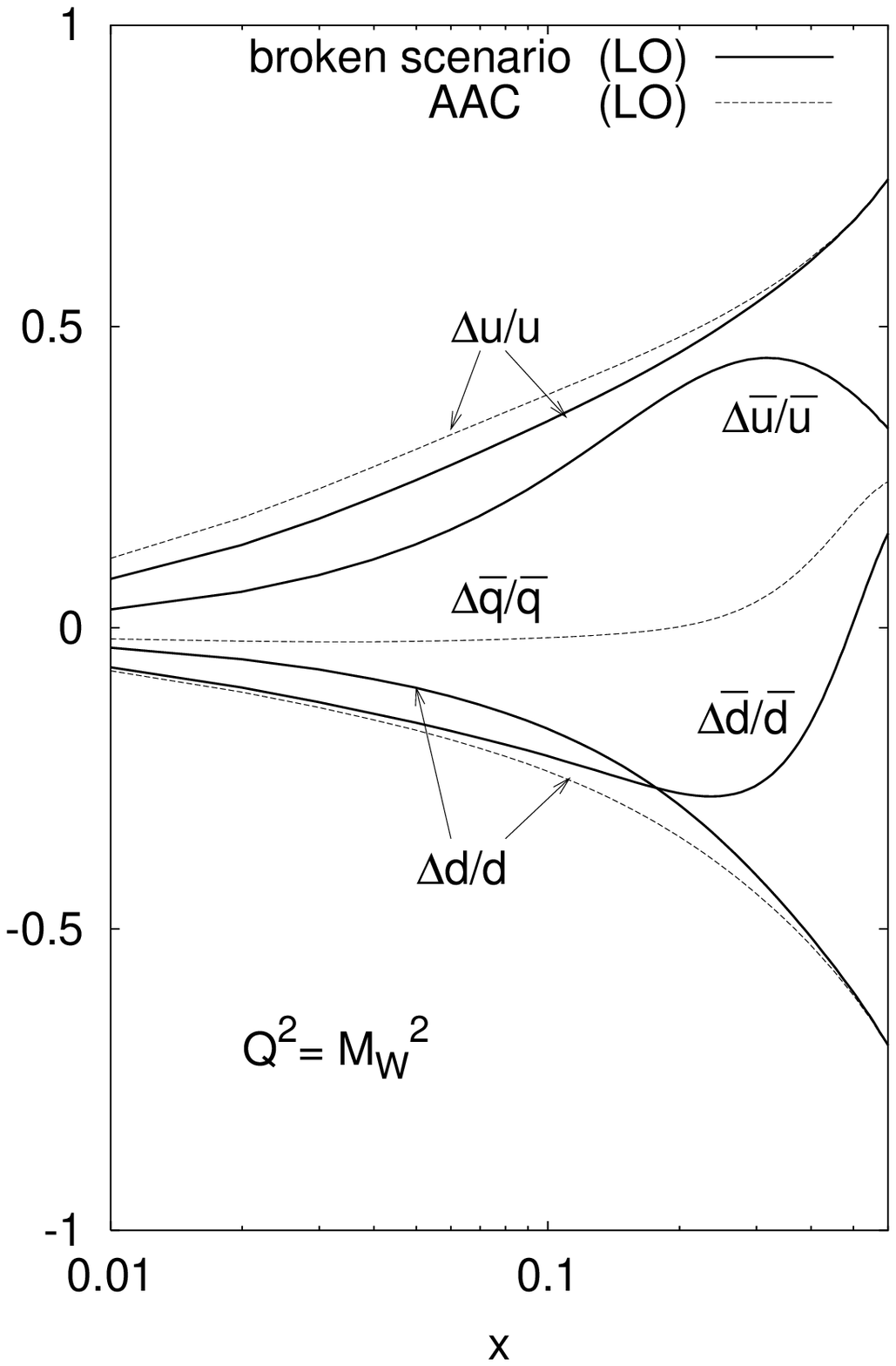,width=8cm,height=11cm,angle=0}
 }}
\hspace*{8cm}
{\large \bf Figure 5}
\end{figure}

\begin{figure}[H]
  \centerline{\hbox{
    \psfig{figure=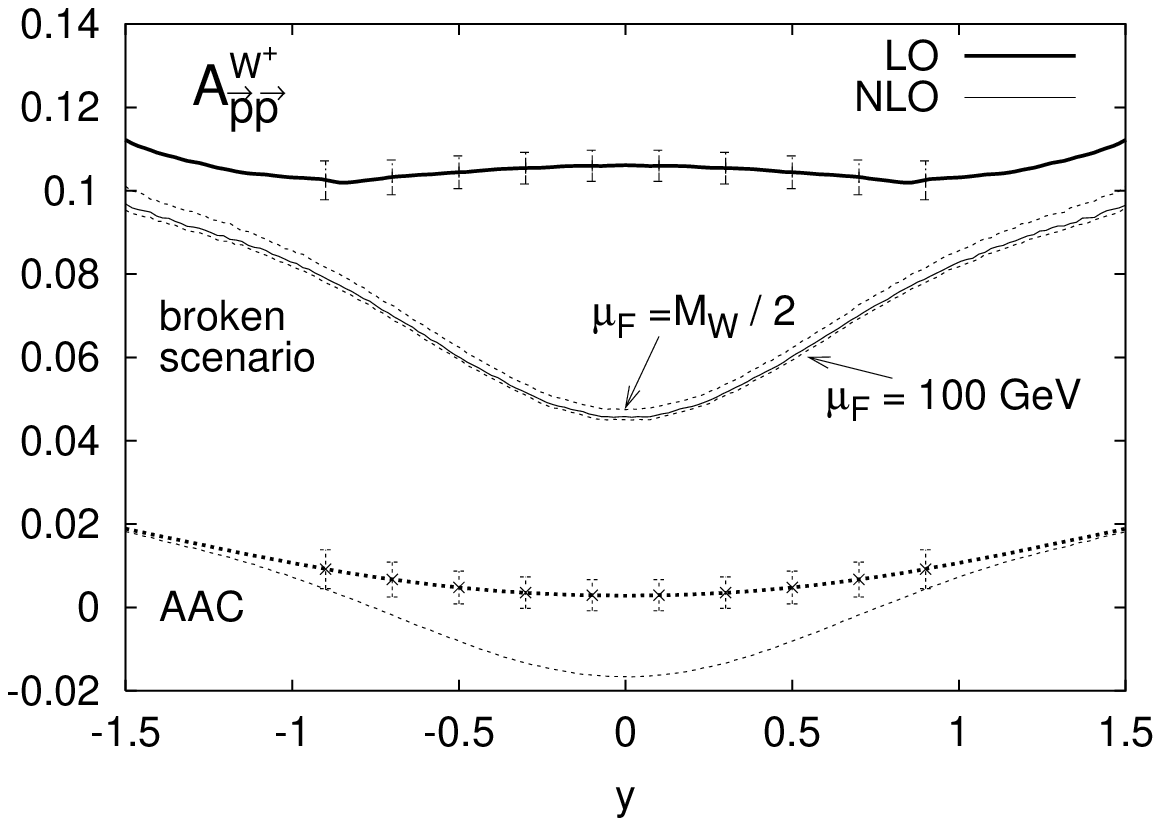,width=9cm,height=6.5cm,angle=0}
     \psfig{figure=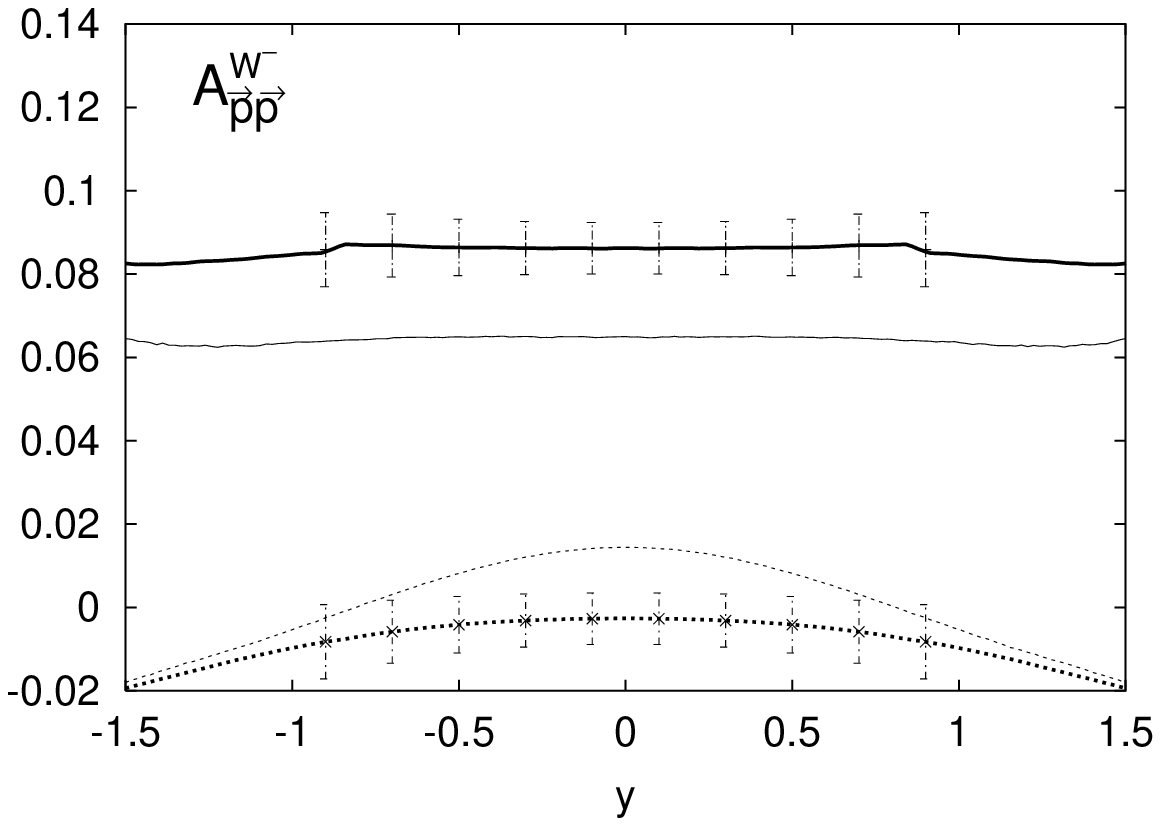,width=9cm,height=6.5cm,angle=0}
 }}
\hspace*{8cm}
{\large \bf Figure 6}
\end{figure}
\vspace*{1cm}
\begin{figure}[H]
  \centerline{\hbox{
    \psfig{figure=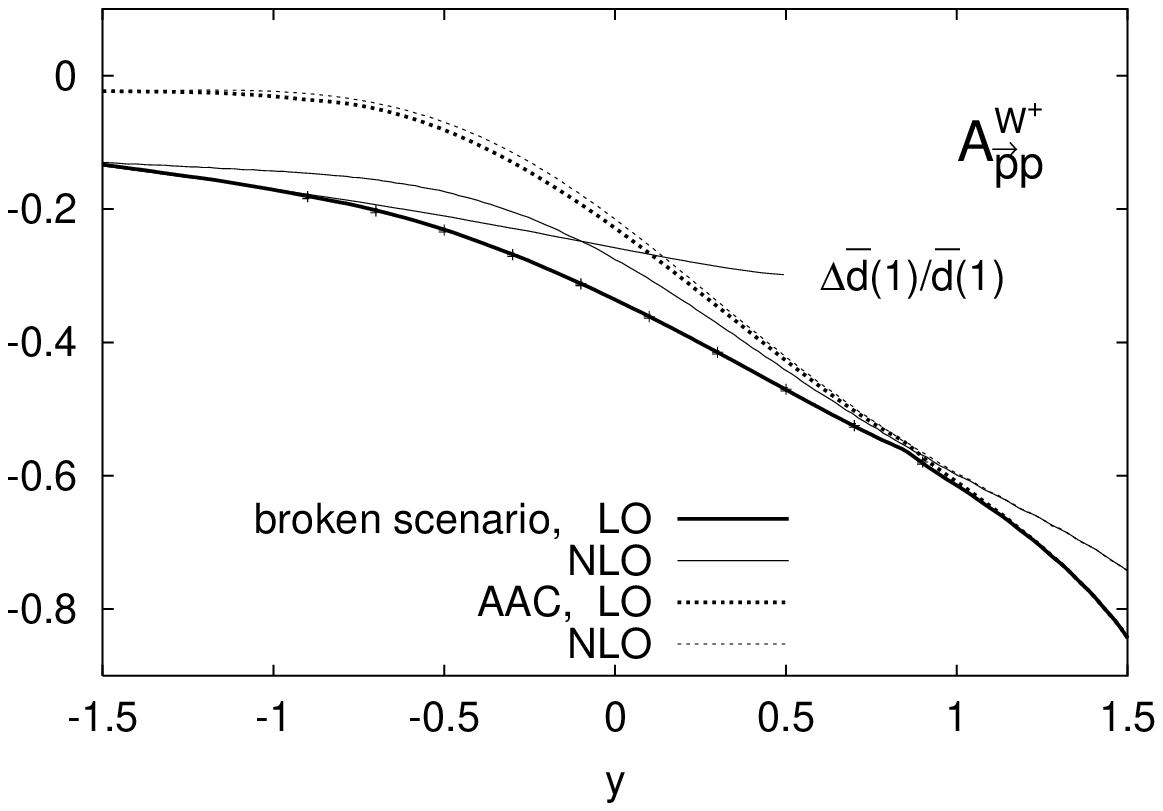,width=9cm,height=6.5cm,angle=0}
     \psfig{figure=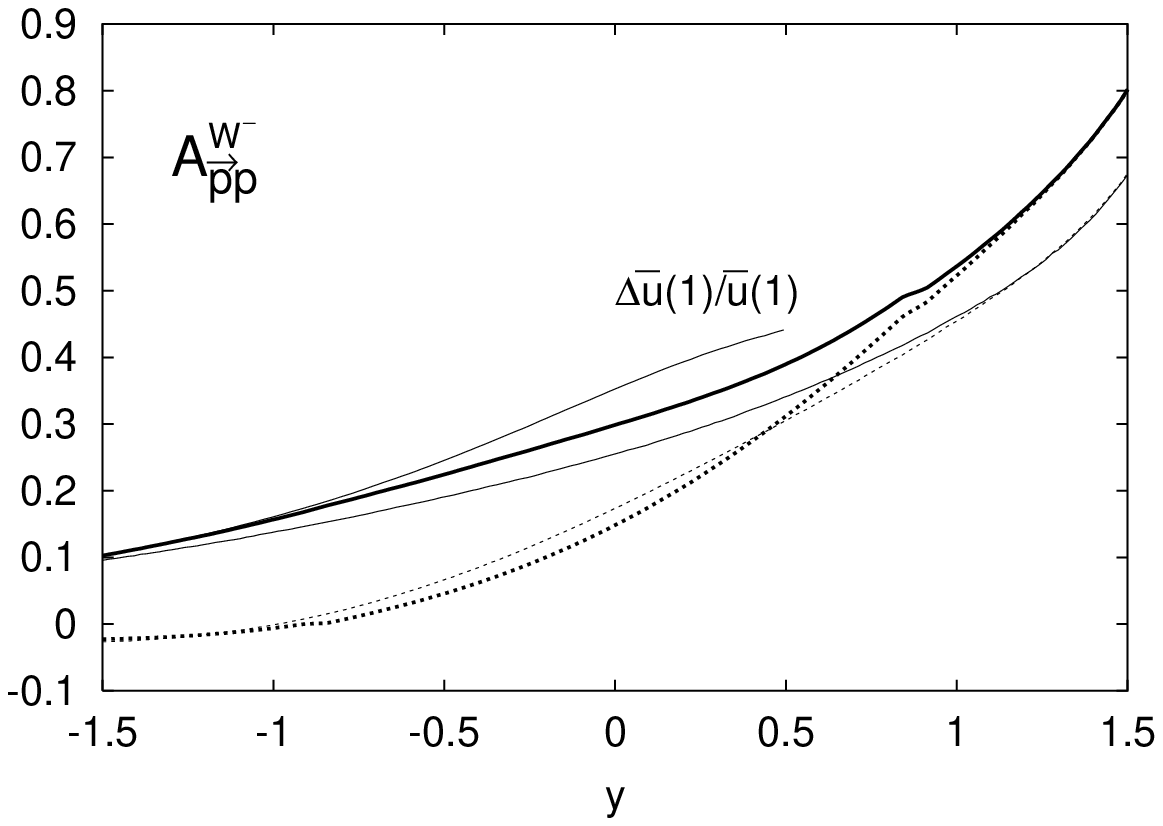,width=9cm,height=6.5cm,angle=0}
 }}
\hspace*{8cm}
{\large \bf Figure 7}
\vspace*{8cm}
\end{figure}

\begin{figure}[H]
  \centerline{\hbox{
    \psfig{figure=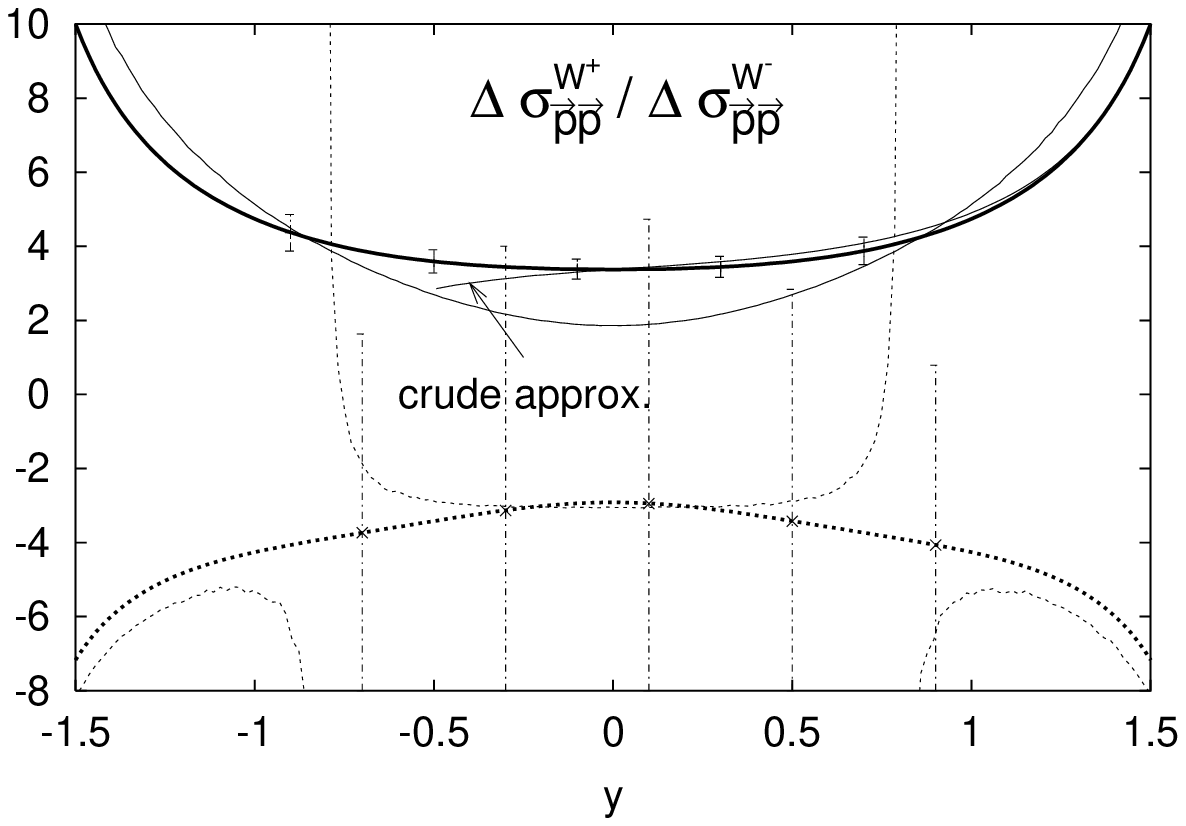,width=9cm,height=6.5cm,angle=0}
     \psfig{figure=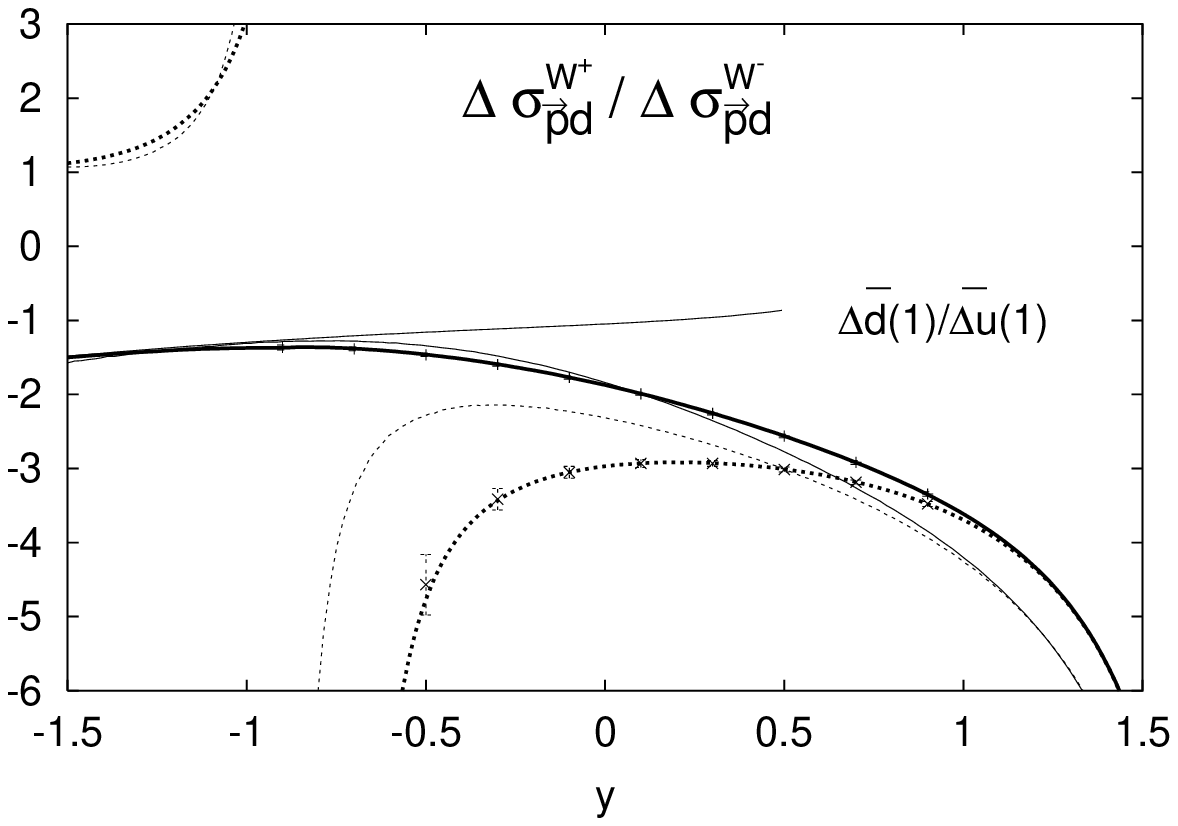,width=9cm,height=6.5cm,angle=0}
 }}
\hspace*{8cm}
{\large \bf Figure 8}
\end{figure}
\begin{figure}[H]
  \centerline{\hbox{
    \psfig{figure=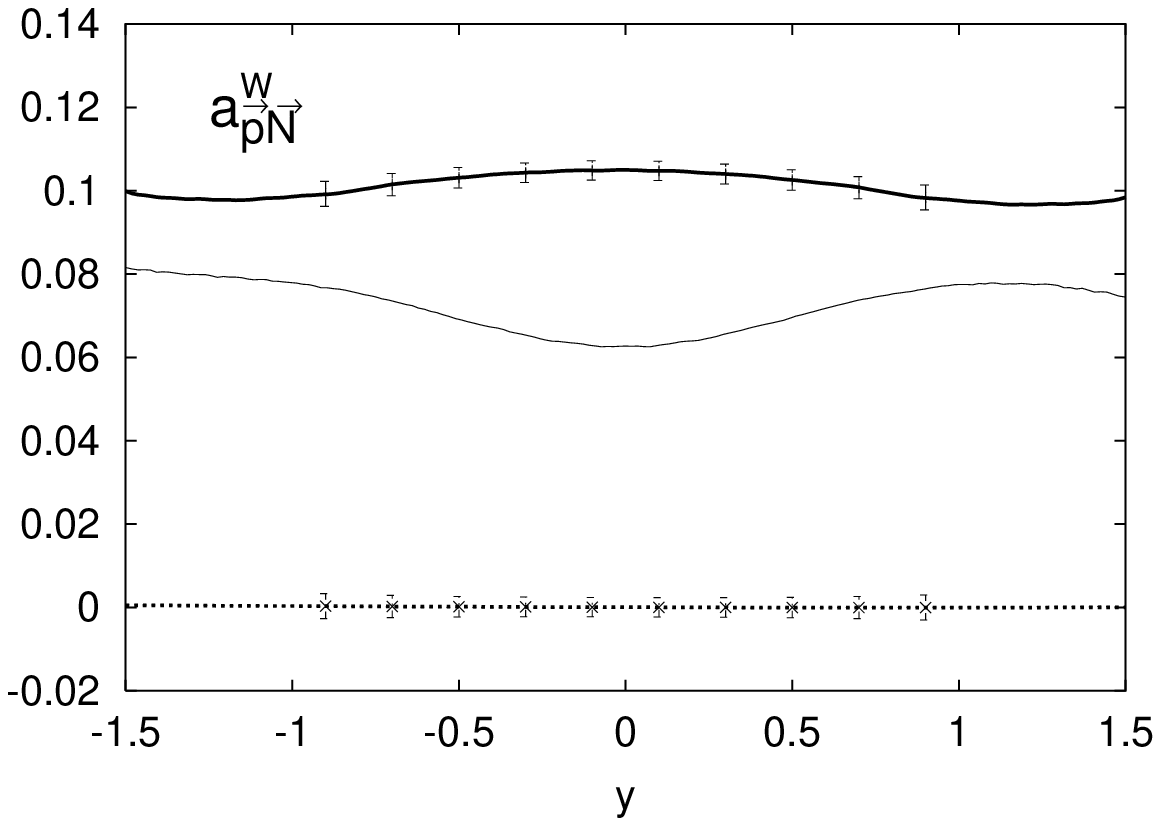,width=9cm,height=6.5cm,angle=0}
     \psfig{figure=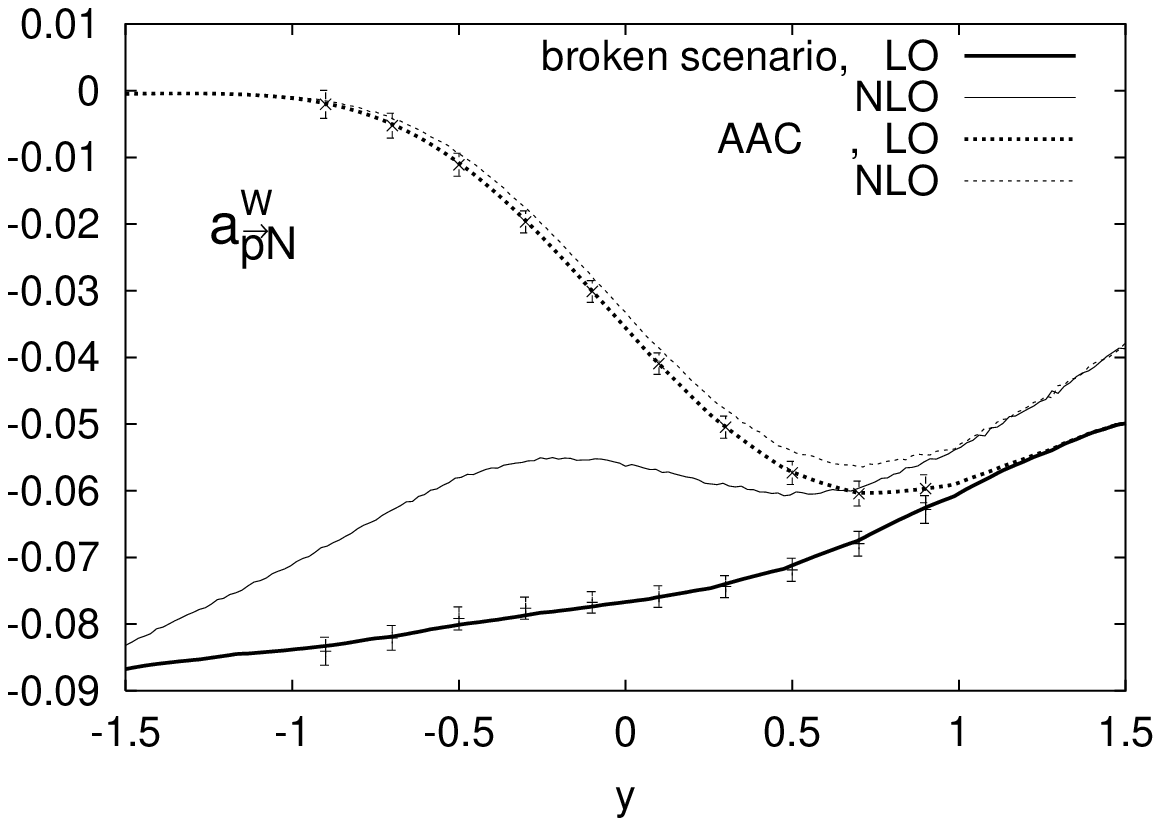,width=9cm,height=6.5cm,angle=0}
 }}
\hspace*{8cm}
{\large \bf Figure 9}
\end{figure}
\begin{figure}[H]
  \centerline{\hbox{
    \psfig{figure=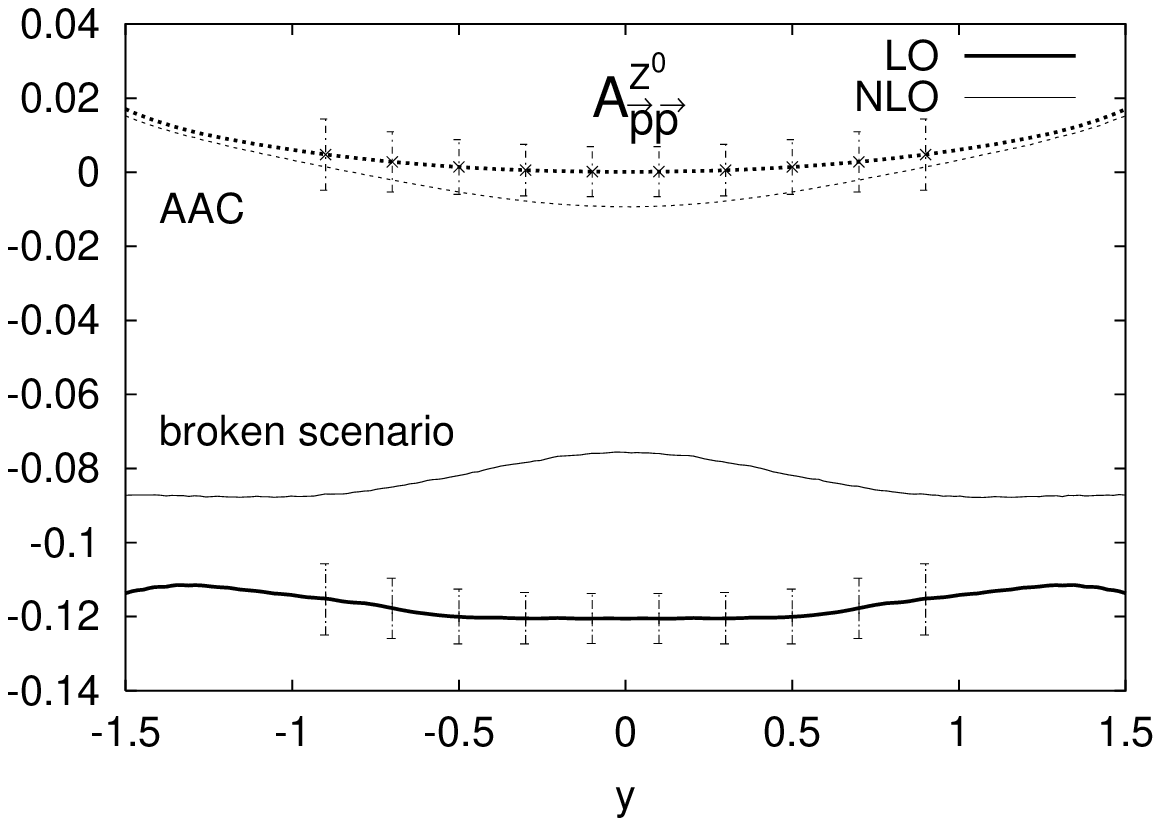,width=9cm,height=6.5cm,angle=0}
     \psfig{figure=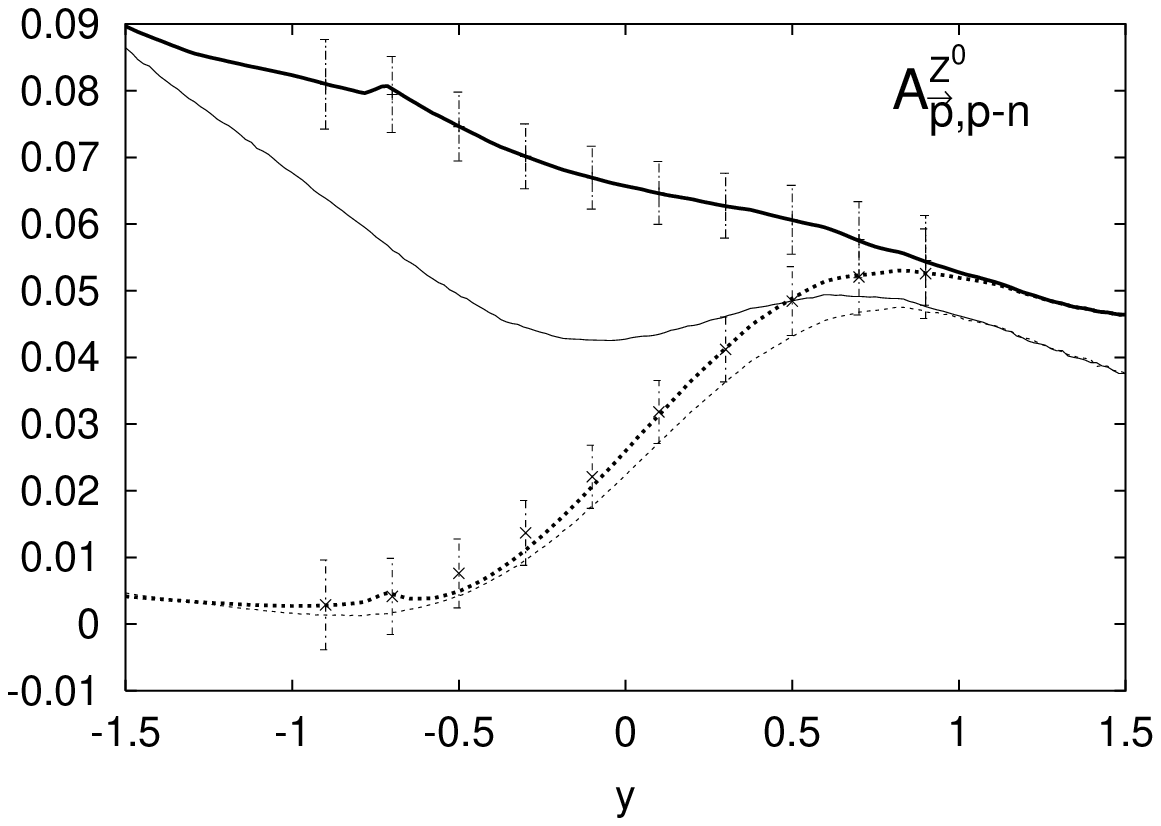,width=9cm,height=6.5cm,angle=0}
 }}
\hspace*{8cm}
{\large \bf Figure 10}
\end{figure}
\end{document}